\documentclass{aims}
\usepackage{amsmath}
\usepackage{array}
\usepackage[T1]{fontenc}
  \usepackage{paralist}
  \usepackage{graphics} 
  \usepackage{epsfig} 
\usepackage{graphicx}  \usepackage{epstopdf}
 \usepackage[colorlinks=true]{hyperref}
 \usepackage{caption}
\usepackage{subcaption}
\hypersetup{urlcolor=blue, citecolor=red}
\newcommand{\Hawaii}{Hawai\kern.05em`\kern.05em\relax i}
\newcommand{\Manoa}{M\=anoa}

\def\currentvolume{X}
 \def\currentissue{X}
  \def\currentyear{200X}
   \def\currentmonth{XX}
    \def\ppages{X--XX}

\newtheorem{theorem}{Theorem}[section]

\newtheorem{lemma}[theorem]{Lemma}

\theoremstyle{definition}
\newtheorem{definition}[theorem]{Definition}
\newtheorem{remark}{Remark}

\newcommand{\ep}{\varepsilon}
\newcommand{\eps}[1]{{#1}_{\varepsilon}}

\usepackage[style=trad-plain,backend=bibtex,sorting=nyt,firstinits=true,
maxbibnames=99,minbibnames=99,maxcitenames=1,natbib=true,
url=false,isbn=false,doi=false,eprint=false]{biblatex}

\DeclareFieldFormat{titlecase}{#1}

\title[A study of compartment and agent based models] 
      {A study of computational and conceptual complexities of compartment and agent based models}

\author[Prateek Kunwar and Oleksandr Markovichenko]{}


 \email{kunwarp@hawaii.edu}
 \email{markov@math.hawaii.edu}
 \email{chyba@hawaii.edu}
 \email{ymileyko@hawaii.edu}
 \email{koniges@hawaii.edu}
 \email{tlee3@hawaii.edu}

\thanks{The first five authors are supported by NSF grant No. 2030789.}

\thanks{$^*$ Corresponding author: chyba@hawaii.edu}

\begin{document}
\maketitle

\centerline{\scshape Prateek Kunwar and Oleksandr Markovichenko}
\medskip
{\footnotesize
 \centerline{Department of Mathematics, University of \Hawaii{} at \Manoa{}}
   \centerline{ Honolulu, HI 96822, USA}
} 

\medskip

\centerline{\scshape Monique Chyba$^*$ and Yuriy Mileyko}
\medskip
{\footnotesize
 \centerline{ Department of Mathematics, University of \Hawaii{} at \Manoa{}}
   \centerline{Honolulu, HI 96822, USA}
}

\centerline{\scshape Alice Koniges}
\medskip
{\footnotesize
 \centerline{ Data Science Institute, University of \Hawaii{} at \Manoa{}}
   \centerline{Honolulu, HI 96822, USA}
}

\centerline{\scshape Thomas Lee}
\medskip
{\footnotesize
 \centerline{ Office of Public Health Studies, University of \Hawaii{} at \Manoa{}}
   \centerline{Honolulu, HI 96822, USA}
}
\bigskip


\begin{abstract}
The ongoing COVID-19 pandemic highlights the essential role of mathematical models in understanding the spread of the virus along with a quantifiable and science-based prediction of the impact of various mitigation measures. Numerous types of models have been employed with various levels of success. This leads to the question of what kind of a mathematical model is most appropriate for a given situation. We consider two widely used types of models: equation-based models (such as standard compartmental epidemiological models) and agent-based models. We assess their performance by modeling the spread of COVID-19 on the Hawaiian island of Oahu under different scenarios. We show that when it comes to information crucial to decision making, both models produce very similar results. At the same time, the two types of models exhibit very different characteristics when considering their computational and conceptual complexity. Consequently, we conclude that choosing the model should be mostly guided by available computational and human resources.
\end{abstract}

\section{Introduction}
SARS-CoV-2 (COVID-19) has impacted not only health, but the economy and how we live daily life.  However, it crept onto the world stage at the end of 2019 in Wuhan, China, where health officials reported a cluster of pneumonia cases of unknown cause.  The first reported case of COVID-19 in the United States came from an asymptomatic male who returned from China on January 15, 2020.  By January 19, Chinese officials closed off travel in and out of Wuhan and on January 30, 2020 the World Health Organization (WHO) declared a global health emergency. 

COVID-19 was officially named on February 11, as it continued to spread across Asia and Europe.  While countries in those regions started to see massive increases in cases, hospitalizations and fatalities during the initial months of the outbreak, the United States did not report its first death until February 29.  By March 26, the United States led the world in confirmed cases.  While many European and Latin American countries were already in full shut down due to COVID-19, the United States Government left the shut down to individual states, resulting in a surge of cases and deaths.  In July 2020, the United States reached 68,000 daily cases for the first time.

During the Fall months, cases started to level off.  However during this period of relative calm, a more serious variant of COVID-19 was mutating and spreading in the United Kingdom.  In November 2020, the B.1.117 COVID-19 variant (UK variant) was detected in the United Kingdom and accounted for 43 percent of all COVID-19 cases by December 2020.  In South Africa, another COVID-19 variant, B.1.351, emerged independently of the UK variant around the same time.  The UK variant is associated with increased transmission and risk of death, while the South African variant shows evidence that it may decrease the neutralization by some monoclonal and polyclonal antibodies.  With the recent surge of cases in India, it was inevitable that India would have its own variants of concern, primarily the B.1.617.  The WHO has added it to its list of variants of concern and England already identified a sub strain of the Indian variant in early May.

The accelerated research, production, and distribution of the COVID-19 vaccines have had an impact on slowing the spread of COVID-19 in the areas that had access to enough supply of the vaccine.  As of June 1, 2021, only 5.5 percent of the world's population have been vaccinated.  While countries such as the United States, Israel, and England have at least 38 percent of their country fully vaccinated, helping to prevent future spread of COVID-19.  Yet many first world countries have yet to even break 10 percent of their population fully vaccinated, including Japan, South Korea, and Canada. While the mRNA vaccines allow for rapid adjustments due to variants, the possibility of a mutation that escapes the current mRNA vaccine remains possible, though unlikely.  This possibility will remain until both industrialized and under-developed countries increase their percentage of population fully vaccinated. 

Mathematicians have found themselves at the front seat of this race against COVID-19. Indeed, modeling is a powerful tool to address key questions such as: when and which non-pharmaceutical mitigation measures should be implemented? how to allocate efficiently vaccines? impact from the new variants? when can we relax travel restrictions? However, there is still a lot of unanswered questions and challenges regarding the outcome of several models as well as their limitations. It is unclear at this time if there is a ``better'' model, and while most of the challenges in epidemiological forecasting come from incomplete data and impossibility to model people's behavior, there is still the question of what model to use when and for what purpose. 

There are primarily two different types of epidemiological models: differential equation-based (EBM) and agent based (ABM). We here focus on two such models: a discrete compartmentalized SEIR model that we developed and the COVID-19 Agent-based Simulator (Covasim) \cite{Kerr2020.05.10.20097469} that we altered to include some specific attributes. 
The first one is deterministic while the second one is stochastic which makes them different in design, however it is expected that their forecasting converge provides similar assumptions since they do model the same pandemic.
It is often stated that EBMs are simpler and faster to compute, while ABMs are more detailed and computationally expensive.\cite{Kerr2020.05.10.20097469} Our goal is to test this assumption concretely on a specific data set that we have extensive knowledge of so we can determine for future studies the benefits and pitfalls of each of these different simulation methods. In particular, we use Honolulu county in the Hawaiian archipelago as a test-bed population for our simulations. We show that rather than treating the two models as two distinct ways to obtain the same results we should exploit advantages of both throughout a pandemic simulation, particularly when the simulation is used in a predictive real-time fashion.  Throughout the current COVID-19 pandemic, most results and forecasting come from one model but not a combination of both. We consider what can be learned from running both models side-by-side; taking and applying the best of each model using the measured data. We also note the conceptual and computational requirements of each of the models for this particular test-bed population.

The outline of the paper is as follows. In section \ref{section-epimodels} we introduce the concepts of compartmental and agent based models and explain how adding features such as travelers and vaccines in the models increases conceptual complexity. We also illustrate our model on data from Honolulu County. Section \ref{section-advlimmodels} discusses conceptual and computational complexity for both models, as well as optimization for data fitting. A benchmark example is provided to analyse performance of both models. We end the paper with a conclusion, section \ref{section-conclusion}.  
\section{Epidemiological Models}
\label{section-epimodels}
Infectious disease modeling has been used by epidemiologists and mathematicians for years, however the COVID-19 pandemic has really highlighted the importance of understanding the purpose and functionality of these models.  At its most core function, the intent of these epidemiological models is to estimate the spread of the infectious disease across a population based on some core epidemiology determinants: incubation period, duration of infectious period, population size, and R0 (the reproductive value).  Our specific implementations of these two models use more variables and complex interactions than other more standard implementations \cite{Cuevas,PREM2020e261,Radulescu, Truszkowska}. 
The purpose of this added functionality is to better account for the nuances and assumptions in a real-world situation.    

The fundamental difference between the two types of models is that EBM captures aggregate behavior over the population while ABM captures agent interactions and progression of the disease over each individual. EBM are typically less computational expensive, and easier to use since they require less information. However, they provide only large-scale information on the spread of the disease compared to ABM models that can take into account spatial information in much more detail. 

\subsection{Compartmental Models}
\begin{figure}[htp!]
\begin{center}
  \includegraphics[width=4.5in]{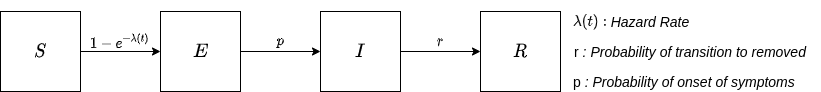}\\
  \caption{Basic SEIR Model diagram and parameters.}\label{Fig_SeirModel}
  \end{center}
\end{figure}

The origin of compartmental models, also called Equation Based Models (EBMs), in the study of infectious diseases is from the works of Ross (1916) \cite{1916RSPSA..92..204R}, Ross and Hudson (1917)\cite{RossHudson}, Kermack and McKendrick in 1927\cite{Kermack} and Kendall (1956) \cite{Kendall}. In those models, the major assumption is that the population is divided into compartments based on the nature of the evolution of the disease. The population size is typically assumed to be fixed and by design is assumed homogeneous within each compartment. The basic model consists of three compartments - Susceptible(S), Infected(I) and Removed(R). Variations of this model may include compartments such as Exposed(E) or a loop back into susceptible in case of diseases with no immunity against re-infection. See Fig.~\ref{Fig_SeirModel}, where the hazard rate $\lambda(t)=\beta\frac{I(t)}{N}$, with $\beta$ denoting the baseline transmission rate and $N$ denoting the (fixed) total population.

 Variants of the SIR model can be differentiated based their treatment of vital dynamics.  Vital dynamics refers to life outcomes such as births and deaths \cite{Hethcote1989}.  In very simple models (or with diseases that are not prolonged) the SIR model without vital dynamics is ideal.  In models that are more detailed, or if the disease is present in the population for a prolonged period of time, the SIR model with vital dynamics is more suitable.  However, the duration of the ``prolonged period of time'' is at least 10 years \cite{Ansumali}.  It is too early to tell where COVID-19 falls regarding the possibility of it being endemic in particular pockets in the world. Historically, the SIR model has been used to estimate the impact of highly infectious diseases, such as smallpox \cite{DelValle}.

Of the various compartmental models, we use one inspired by \cite{LloydSmith2003}, which is based on a standard discrete and deterministic SEIR model. Some classical SEIR models include Cooke and Driessche \cite{Cooke}, who introduced the SEIR model with two delays and Li et al. \cite{Li} studied global dynamics with both non-linear and standard incidence rates. We assume a given population is divided into four compartments: Susceptible (not currently infected), Exposed (infected with no symptoms), Infected (infected with symptoms), and Removed (recovered or deceased). We subdivide the entire population into two additional groups: the general community (C), healthcare workers (H). This is motivated by the fact that healthcare workers are potentially more exposed to a virus but also use better protection, and therefore should interact differently with the community during a pandemic. In addition, during the severe acute respiratory syndrome (SARS) epidemic in 2003, healthcare workers formed a large fraction of the infected population \cite{LloydSmith2003}. This has also been which suggested to be the case for COVID-19 \cite{NGUYEN2020e475,Sabetian}. 

\begin{figure}[htp!]
\begin{center}
  \includegraphics[width=4.5in]{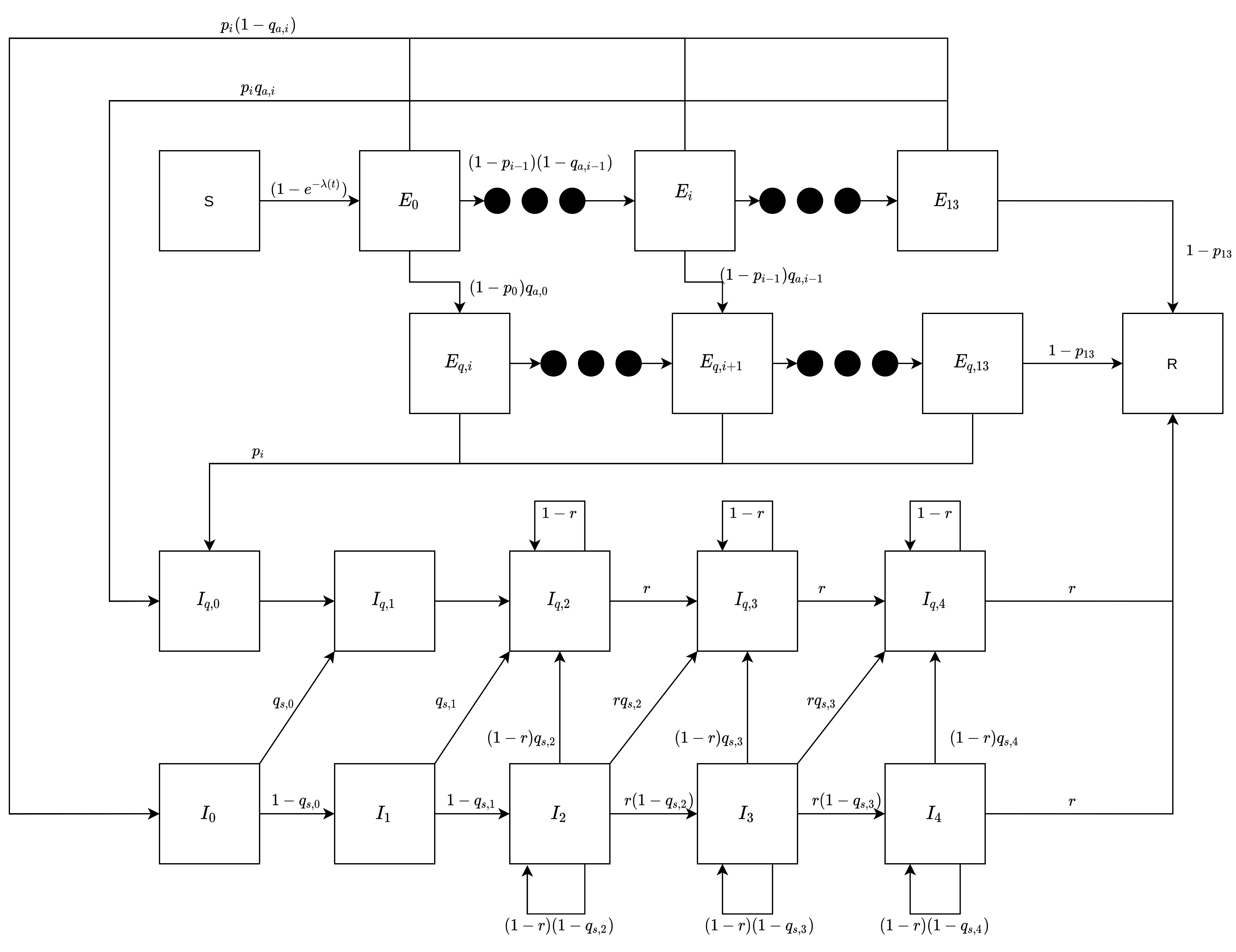}\\
  \caption{Diagram of our basic compartmental  model.}\label{Fig_Seir_Flow_2_base_trans}
  \end{center}
\end{figure}

In our model, Exposed and Infected (in each population group) are split into multiple stages per day to better reflect the progression of the disease. Individuals in isolation (including hospitalization) are similarly distinguished. The dynamics of the two population groups are essentially the same and are represented using the diagram in Fig.~\ref{Fig_Seir_Flow_2_base_trans} with variables described in Tab.~\ref{Table_Seir_Flow_2_base_trans}. 

\begin{table}[h!]
\centering
\begin{tabular}{ |p{3cm}||p{8cm}| }
 \hline
 \multicolumn{2}{|c|}{Description of the variables in the EBM model.} \\
 \hline
 Variable     & Description \\
 \hline
 \hline
 $S(t)$   & Number of total susceptible individuals    \\
 \hline
$E_{i}(t)$ (resp. $E_{q,i}(t)$)   & Number of asymptomatic infected individuals $i$ days after exposure who are not quarantined (resp. qurantined)   \\
 \hline
 $I_{j}(t)$ (resp. $I_{q,j}(t)$), $i=0,1$   & Number of symptomatic infected individuals $i$ days after the onset of symptoms who are not isolated (resp. isolated) \\
 \hline
 $I_{j}(t)$ (resp. $I_{q,j}(t)$), $j=3,4,5$   & Number of symptomatic infected individuals at the nominal stage $i$ of the illness that has not been isolated (resp. isolated). Note that a person can stay at a given stage for several days    \\
 \hline
   $R(t)$ & Number of removed (recovered or deceased) individuals   \\
 \hline
\end{tabular}
\caption{SEIR basic model variables. Isolated accounts for quarantine and hospitalization.}
\label{Table_Seir_Flow_2_base_trans}
\end{table}
 By choosing the probabilities of moving from one stage to the other, we are able to model the observation  (according to the Centers for Disease Control and Prevention (CDC) as well as other sources) that about 40\% of people who contract SARS-CoV-2 remain asymptomatic and that the incubation period for those who do develop symptoms is between 2 to 14 days after exposure, with the mean period being between 4 and 6 days\cite{Park}. Those who do not develop symptoms after 14 days are moved directly from the Exposed compartment to the Removed compartment. The quarantine sub-compartment $E_{q,i}$ is also broken down into 14 stages and is used to model the effect of contact tracing and the reduced transmission rate for quarantined individuals. The infected individuals go through 5 stages, of which the first two represent the first two days of being symptomatic whereas the last three represent the phase where the immune system is fighting the disease. Since the last three stages can go on for more than one day each, there is a variability in the number of days any given person can spend at each stage. Our model assumes that the symptomatic phase of the illness lasts at least 5 days. This can be seen in Figure \ref{Fig_Seir_Flow_2_base_trans} as the feedback loops in $I_2,I_3$ and $I_4$ where only $20\%$ of people in each of these compartments move on to the next one. The remaining $80\%$ stay in the same compartment for the next iteration. 
The parameters in Fig.~\ref{Fig_Seir_Flow_2_base_trans} are described in Tab.~\ref{tab:parameters-common}. The parameter $\beta$, the basal transmission rate, is optimized to fit the data. Addition of compartments, mask mandate, contact tracing, travelers and vaccinations modify $\beta$ according to equations (\ref{basic_c_lambda}), (\ref{basic_h_lambda}), (\ref{visitor_c_lambda}), (\ref{visitor_v_lambda}), (\ref{vaccine_c_lambda}) and (\ref{vaccine_h_lambda}) to obtain the hazard rate $\lambda_i$, $(i=c,h,v)$ which governs the dynamics of the evolution of disease for each category. We introduce parameters $p_i$ for the probability to develop symptoms on day $i$, and refine them such that if symptoms do develop, it takes between 2 to 14 days, with a mean between 4 and 6 days\cite{Park}, while assuming that about 40\% of all infections remain asymptomatic. The values of $q_{s,i}$ are chosen to reflect the prediction that symptomatic individuals are likely to quarantine, with a probability of 0.1, 0.4, 0.8, 0.9 and 0.99 for the first five days of symptoms. For healthcare population these probabilities are assumed to be slightly higher at 0.2, 0.5, 0.9, 0.98 and 0.99 for the first five days of symptoms. In addition, the parameter $r$ is the probability of transitioning from one stage of the illness to the next (with the final stage being recovery or death). Based on prior work \cite{GlopbalHealth2}, we chose $r$ to yield an expected length of illness of 17 days. 

The equations for the dynamics are given in (\ref{basic_S})-(\ref{basic_R}). 
\begin{align}
S(t+1) &= e^{-\lambda(t)}S(t)\label{basic_S}\\
E_0(t+1) &= (1-e^{-\lambda(t)})S(t)\\
E_{i}(t+1)& = (1-p_{i-1})(1-q_{a,i-1})E_{i-1}(t),\quad i=1,\ldots,13\\
E_{q,i}(t+1)& = (1-p_{i-1})(q_{a,i-1}E_{i-1}(t)+E_{q,i-1}(t)), \quad i=1,\ldots,13\\
I_{0}(t+1) & =\sum_{i=0}^{13} p_i(1-q_{a,i})E_{i}(t) \\
I_{1}(t+1) & =(1-q_{s,0})I_{0}(t)\\
I_{2}(t+1) & =(1-q_{s,1})I_{1}(t)+(1-r)(1-q_{s,2})I_{2}(t)\\
I_{j}(t+1) & =r(1-q_{s,j-1})I_{j-1}(t)+(1-r)(1-q_{s,j})I_{j}(t),\quad j=3,4\\
I_{q,0}(t+1) & =\sum_{i=0}^{13} p_i(q_{a,i}E_{i}(t)+E_{q,i}(t))\\
I_{q,1}(t+1) & = I_{q,0}(t)+q_{s,0}I_{0}(t)\\
I_{q,2}(t+1) & = I_{q,1}(t)+q_{s,1}I_{1}(t)+(1-r)(q_{s,2}I_{2}(t)+I_{q,2}(t))\\
I_{q,j}(t+1) & = r(q_{s,j-1}I_{j-1}(t)+I_{q, j-1}(t))+(1-r)(q_{s,j}I_{j}(t)+I_{q,j}(t)),\quad j=3,4\\
R(t+1) &= R(t) + rI_{4}(t)+rI_{q,4}(t) + (1-p_{13})E_{13}(t) + (1-p_{13})E_{q,13}(t) \label{basic_R}
\end{align}
As we mentioned, a crucial part of the dynamics relates to the hazard rate. For the general community, group C, we have
\begin{multline}
\lambda_c(t) = \beta(1-p_{mp}(1-p_{me}))\Big[
  (I_c+\varepsilon E_c)+
  \gamma((1-\nu)I_{c,q}+\varepsilon E_{c,q})+\\
  \rho[(I_h+\varepsilon E_h)+  \gamma((1-\nu)I_{h,q}+\varepsilon E_{h,q})]]\Big]/N_c, \label{basic_c_lambda}
\end{multline}
where we suppressed the dependency on $t$ on the right for convenience. We use sub-indices $c$ (community), and $h$ (healthcare workers) to indicate the appropriate group. The subscript $q$ indicates quarantined and isolated individuals. Here $p_{me}$ and $p_{mp}$ represent mask efficiency and mask compliance, respectively. Mask efficiency is chosen to reflect a reduction in transmission of $75\%$. $N_c$ denotes the mixing pool for the general community, computed as
\begin{equation}
    N_c(t) = S_c+E_c+I_c+R_c+\rho(S_h+E_h+I_h+R_h).
\end{equation}
where variables $E$ and $I$ here represent the sum over all the stages within these compartments. 
For the healthcare worker group, we have
\begin{equation}
    \lambda_h(t) = \rho\lambda_c+\beta\eta\Big[
  (I_h+\varepsilon E_h)+
  \kappa\nu(I_{h,q}+I_{c,q})\Big]/N_h, \label{basic_h_lambda}
\end{equation}
where $N_h(t) = S_h+E_h+I_h+R_h$. 

\begin{figure}[htp!]
\begin{center}
  \includegraphics[width=5.2in]{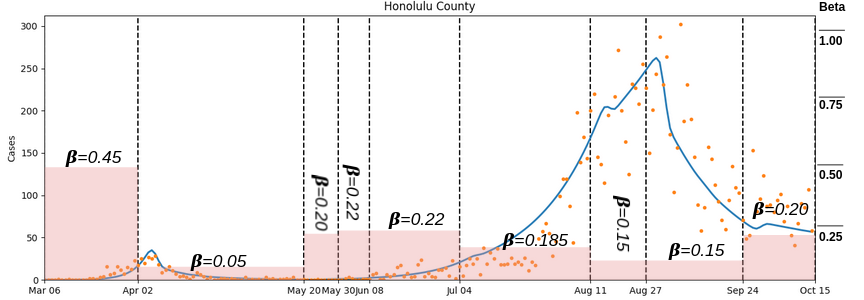}\\
  \caption{Blue denotes the SEIR model fit for Honolulu county from March 6 to October 15, 2020. The dots represent the actual daily cases from the \Hawaii{} DOH dashboard \cite{HawaiiDOH} Included are the optimized transmission rates (aligned with non pharmaceutical mitigation measures that were taken by the State of \Hawaii{}).}\label{Fig_Seir_EBM_fit_Basic}
  \end{center}
\end{figure}

\begin{table}[h!]
\caption{Parameters intrinsic to COVID-19}
\centering
\begin{tabular}{ |>{\raggedright\let\newline\\\arraybackslash\hspace{0pt}}p{6.5cm}||>{\raggedright\let\newline\\\arraybackslash\hspace{0pt}}p{6cm}|}
 \hline
\centerline {Parameter, meaning} & \centerline {Value} \\
 \hline
  \hline
$\beta$, basal transmission rates   &  optimized to fit data  \\
 \hline
  \multicolumn{2}{|c|}{Factors modifying transmission rate}\\
 \hline
$\varepsilon$, asymptomatic transmission (25\% reduction in transmission) & 0.75 \\
 \hline
$\rho$, reduced healthcare worker interactions &  0.8 \\
  \hline
     $\gamma$, quarantine (80\% reduction in transmission)   &  0.2   \\
   \hline
   $\kappa$, hospital precautions & 0.5 \\
   \hline
   $\eta$, healthcare worker precautions & $0.2375$ \\
   \hline
  \multicolumn{2}{|c|}{Population fractions}\\
   \hline
  $p_i$, $i=$0,\ldots,13, onset of symptoms after day $i$  & 0.000792, 0.00198, 0.1056, 0.198, 0.2376, 0.0858, 0.0528, 0.0462,
0.0396, 0.0264, 0.0198, 0.0198, 0.0198, 0\\
  \hline
  $q_{s,i}$, $i=$0,\ldots,4, symptomatic quarantine after day/stage $i$ & C: 0.1, 0.4, 0.8, 0.9, 0.99;\newline H: 0.2, 0.5, 0.9, 0.98, 0.99\\
  \hline
  r, transition to next symptomatic day/stage & 0.2\\
   \hline
   $\nu$, symptomatic hospitalization & 0.11 \\
  \hline
  
\hline
\end{tabular}
\label{tab:parameters-common}
\end{table}
Figure \ref{Fig_Seir_EBM_fit_Basic} represents the optimized fit obtained from the SEIR model for Honolulu County. 

\begin{figure}[htp!]
\begin{center}
  \includegraphics[width=2.5in]{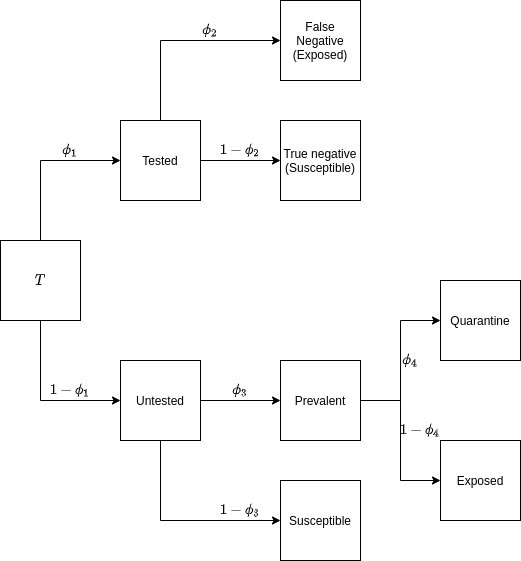}\\
  \caption{Sub-Diagram for travelers assumptions.}\label{traveler_sub_diagram}
  \end{center}
\end{figure}

On October 15, 2020, the State of \Hawaii{} implemented the Safe travel program which brought back tourists to the islands as well as allowed more residents to travel to the mainland. Travelers are an important component of spread of a disease, especially for more isolated locations such as islands or archipelagos (among which for instance New Zealand, Iceland, Japan and Polynesia). To implement travelers, we introduce daily travelers (T) and consider two broad categories of travelers - tourists and returning residents. The returning residents are assumed to behave similar to the existing community members whereas the tourists are assumed to have different behaviour and form a new group (V). For tourists, we assume a 25\% higher basal transmission rate to account for their risk-taking behaviour. We also assume a 50\% reduced interaction of tourists with the community. This is reflected in the parameter $\rho_v$ which can be seen in the expression for hazard rate, $\lambda_c$. To account for some form of safe travel protocol for every region, we make further assumptions about the testing rate
$\phi_{1}$, false negative rate $\phi_{2}$, prevalence $\phi_{3}$ and fraction of untested travelers quarantining $\phi_4$. See Diagram \ref{traveler_sub_diagram} for a visual of our assumptions. 

From these assumptions we compute the coefficients for the number of arriving travelers distributed in each compartment as $v_1, v_2, v_3$:
\begin{eqnarray}
    v_1 =& \phi_1(1-\phi_2) + (1-\phi_1)(1-\phi_3) \\
    v_2 =& (1-\phi_1)\phi_3(1-\phi_4) + \phi_1 \phi_2\\
    v_3 =& (1-\phi_1)\phi_3\phi_4\\
\end{eqnarray}
For simplicity we assume the untested unexposed travelers go directly into the susceptible group even though they might quarantine for 10 days. This is to avoid introducing a new variable of susceptible quarantine individuals.  
\begin{figure}[htp!]
\begin{center}
  \includegraphics[width=5in]{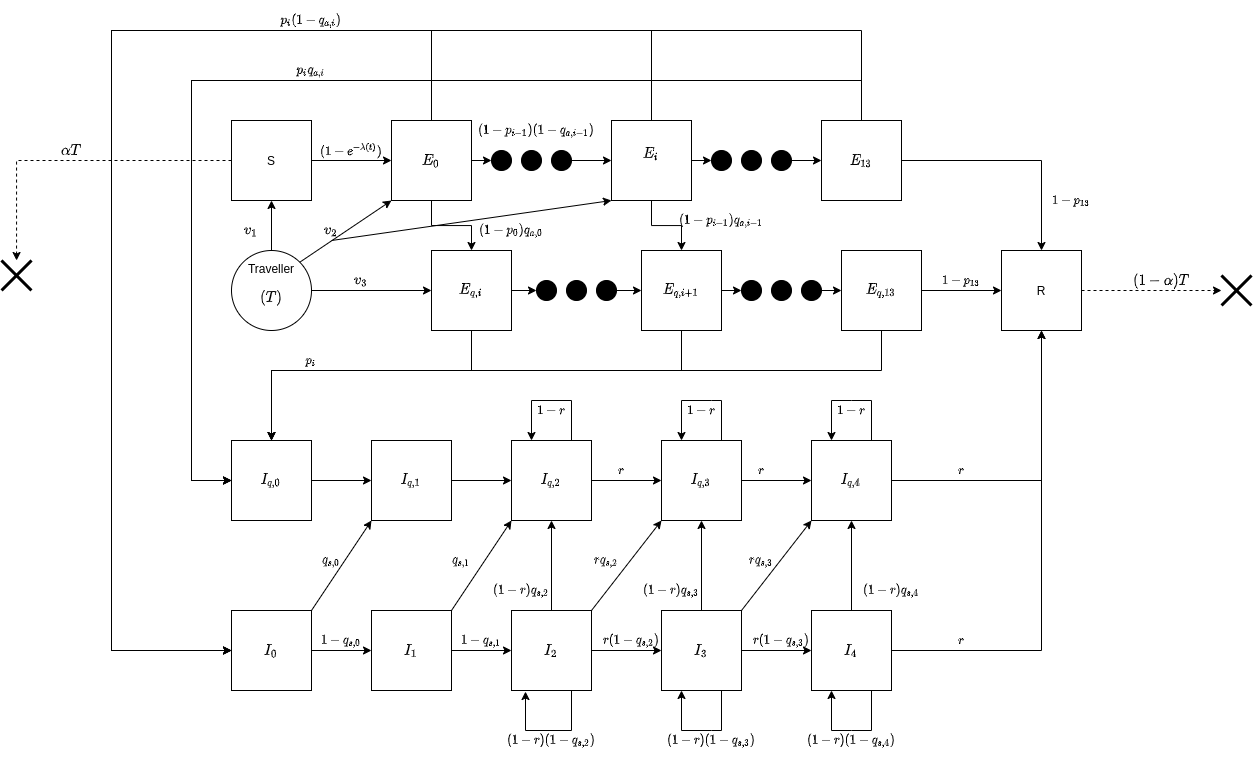}\\
  \caption{Diagram of our compartmental  model with travelers.}\label{Fig_Seir_Flow_2_base_trans_travel}
  \end{center}
\end{figure}

The dynamics equations become
\begin{align}
S(t+1) &= e^{-\lambda(t)}S(t){\color{red} +v_1T}\\
E_0(t+1) &= (1-e^{-\lambda(t)})S(t){\color{red} +\frac{v_2}{5}T}\\
E_i(t+1) &= (1-p_{i-1})(1-q_{a,i-1})E_{i-1}(t){\color{red} +\frac{v_2}{5}T,\quad i=1,\ldots,4}\\
E_{q,i}(t+1)& = (1-p_{i-1})(q_{a,i-1}E_{i-1}(t)+E_{q,i-1}(t)){\color{red} + \frac{v_3}{5}T, \quad i=1,\ldots,5}
\end{align}
where the terms in red account for travelers and $T$ is the average number of travelers entering the region per day. With these assumptions and changes in the model, the equation for the hazard rates get modified to: 
\begin{multline}
\lambda_c(t) = \beta(1-p_{mp}(1-p_{me}))\Big[
  (I_c+\varepsilon E_c)+
  \gamma((1-\nu)I_{c,q}+\varepsilon E_{c,q})+\\
  \rho[(I_h+\varepsilon E_h)+  \gamma((1-\nu)I_{h,q}+\varepsilon E_{h,q})]+\\
  {\color{red} \rho_v[(I_v+\varepsilon E_v)+  \gamma((1-\nu)I_{v,q}+\varepsilon E_{v,q})]]\Big]/N_c},\label{visitor_c_lambda}
\end{multline}
where:
\begin{equation}
    N_c(t) = S_c+E_c+I_c+R_c+\rho(S_h+E_h+I_h+R_h){\color{red}+\rho_v(S_v+E_v+I_v+R_v)}.
\end{equation}
The expression for $\lambda_h$ remains unchanged with the introduction of travelers, however, we now have a new hazard rate, $\lambda_v$ that governs the dynamic of the tourist group (V):
\begin{multline}
\lambda_v(t) = \frac{
  \rho_v N_{c} \lambda_c+
  \beta_v(1-p_{m_p}(1-p_{m_e}))[(I_v+\varepsilon E_v)+  \gamma((1-\nu)I_{v,q}+\varepsilon E_{v,q})]}{\rho_v N_c+N_v}, \label{visitor_v_lambda}
\end{multline}

\begin{table}[h!]
\caption{Parameters intrinsic to travelers }
\centering
\begin{tabular}{ |>{\raggedright\let\newline\\\arraybackslash\hspace{0pt}}p{6.5cm}||>{\raggedright\let\newline\\\arraybackslash\hspace{0pt}}p{6cm}|}
 \hline
\centerline {Parameter, meaning} & \centerline {Value} \\
 \hline
 \hline
  \multicolumn{2}{|c|}{Factors modifying transmission rate}\\
 \hline
$\rho_v$, reduced interaction of travelers with community & 0.5 \\
\hline
$\phi_1$, percentage of tested travelers &  vary by destination (0.86 for Honolulu)\\
\hline
$\phi_2$, false negative test & 0.005 (assuming Nucleic Acid Amplification Test) \\
\hline
$\phi_3$, prevalence &  0.05\\
\hline
$\phi_4$, untested, exposed into quarantine &  0.99\\
  \hline
  
\hline
\end{tabular}
\label{tab:parameters-travelers}
\end{table}

Since the travelers add to the susceptible population, we have to make certain assumptions on how and when they are removed from the model once they leave the region. The number of susceptible travelers is given by $v_1e^{-\lambda(t)}T$, 
and we remove a fraction of them to reflect them leaving the region (represented by the $\alpha$ in Fig.~\ref{Fig_Seir_Flow_2_base_trans_travel}). Tourists are removed from the R compartment proportionally to incoming tourists that got exposed. 

\begin{figure}[htp!]
\begin{center}
  \includegraphics[width=5in]{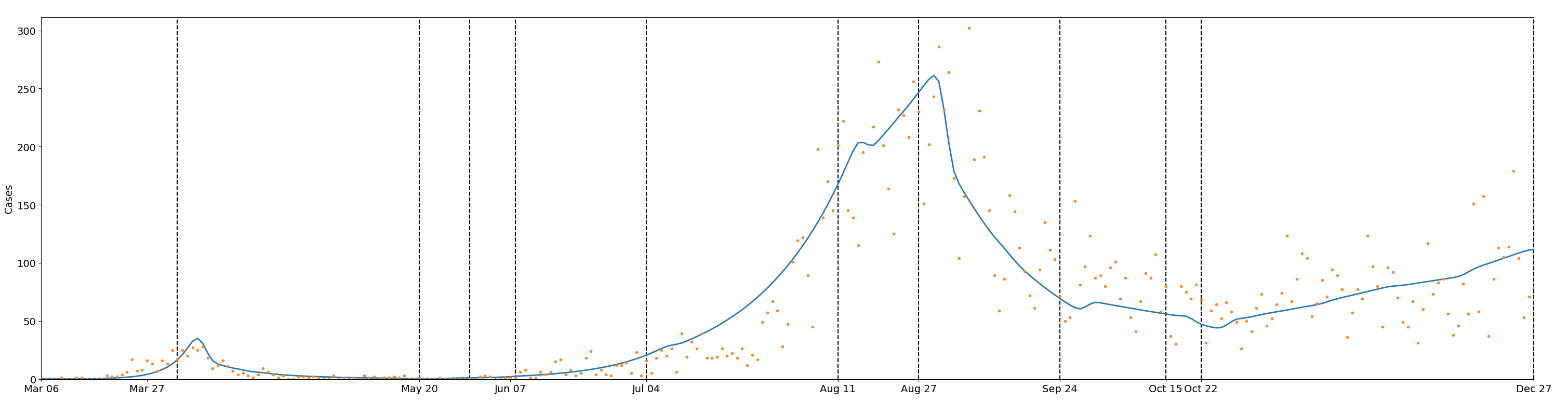}\\
  \includegraphics[width=5in]{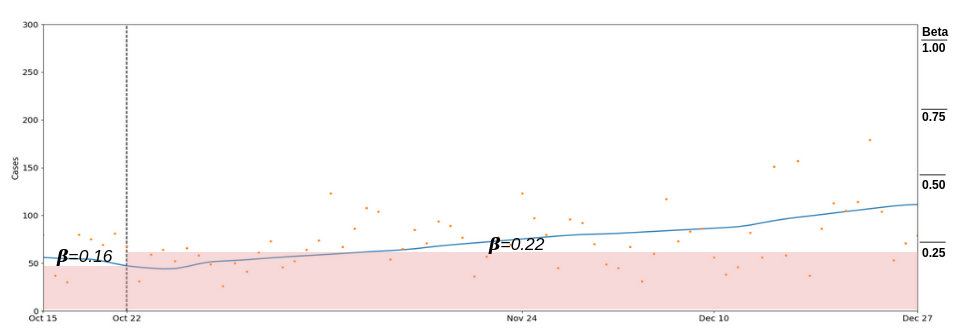}
  \caption{Top: Honolulu County fit from March 6, 2020 including travelers from October 15 to December 27, 2020. Bottom: Honolulu County fit zoomed in for period of October 15 - December 27 with the value for the basal transmission rate. 
  }\label{Fig_Seir_EBM_fit_Travelers}
  \end{center}
\end{figure}

Figure \ref{Fig_Seir_EBM_fit_Travelers} represents the optimized fit obtained from the SEIR model for Honolulu County including the period from October 15 to December 27, 2020 when travelers were re-introduced through the safe travel program. The parameters in Table \ref{tab:parameters-travelers} were used for Honolulu county to find $v_1,v_2,v_3$ as described earlier. The per day average influx of travelers and returning residents are modeled as piece-wise linear functions over two week intervals from October 15, 2020 to April 25, 2021. The data is obtained from \cite{HawaiiSafeTravels} and the values are shown in Table \ref{tab:data-traveler}. 
\begin{table}[h!]
   \captionsetup{justification=centering}
    \caption {Traveler Data for Honolulu County}
    \centering
    \begin{tabular}{|p{3.0cm}|p{3.7cm}|p{4cm}|}
        \hline
         Dates & Average Tourists per day  & Average Returning Residents per day \\
        \hline
        \hline
        Oct 15 - Oct 28 & 1353 & 692 \\
        \hline
        Oct 29 - Nov 11 &2124 &716 \\
        \hline
        Nov 12 - Nov 25 &3051 &967 \\
        \hline
        Nov 26 - Dec 9 &2028 & 951\\
        \hline
        Dec 10 - Dec 23 &4724 & 1014 \\
        \hline
        Dec 24 - Jan 6 &2195 & 1018\\
        \hline
        Jan 7 - Jan 20 & 1522 & 1053\\
        \hline
        Jan 21 - Feb 3 & 1531 & 710\\
        \hline
        Feb 4 - Feb 18 & 2828 & 843\\
        \hline
        Feb 19 - Mar 4 & 2832 & 942\\
        \hline
        Mar 5 - Mar 19 & 4483 & 1017\\
        \hline
        Mar 20 - Apr 5 & 6263 & 1543\\
        \hline
        Apr 6 - Apr 20 & 6231 & 1087\\
        \hline
        Apr 21 - Apr 25 & 5683 & 1331\\
        \hline
        
    \end{tabular}
    \label{tab:data-traveler}
\end{table}

\begin{figure}[htp!]
\begin{center}
  \includegraphics[width=4.5in]{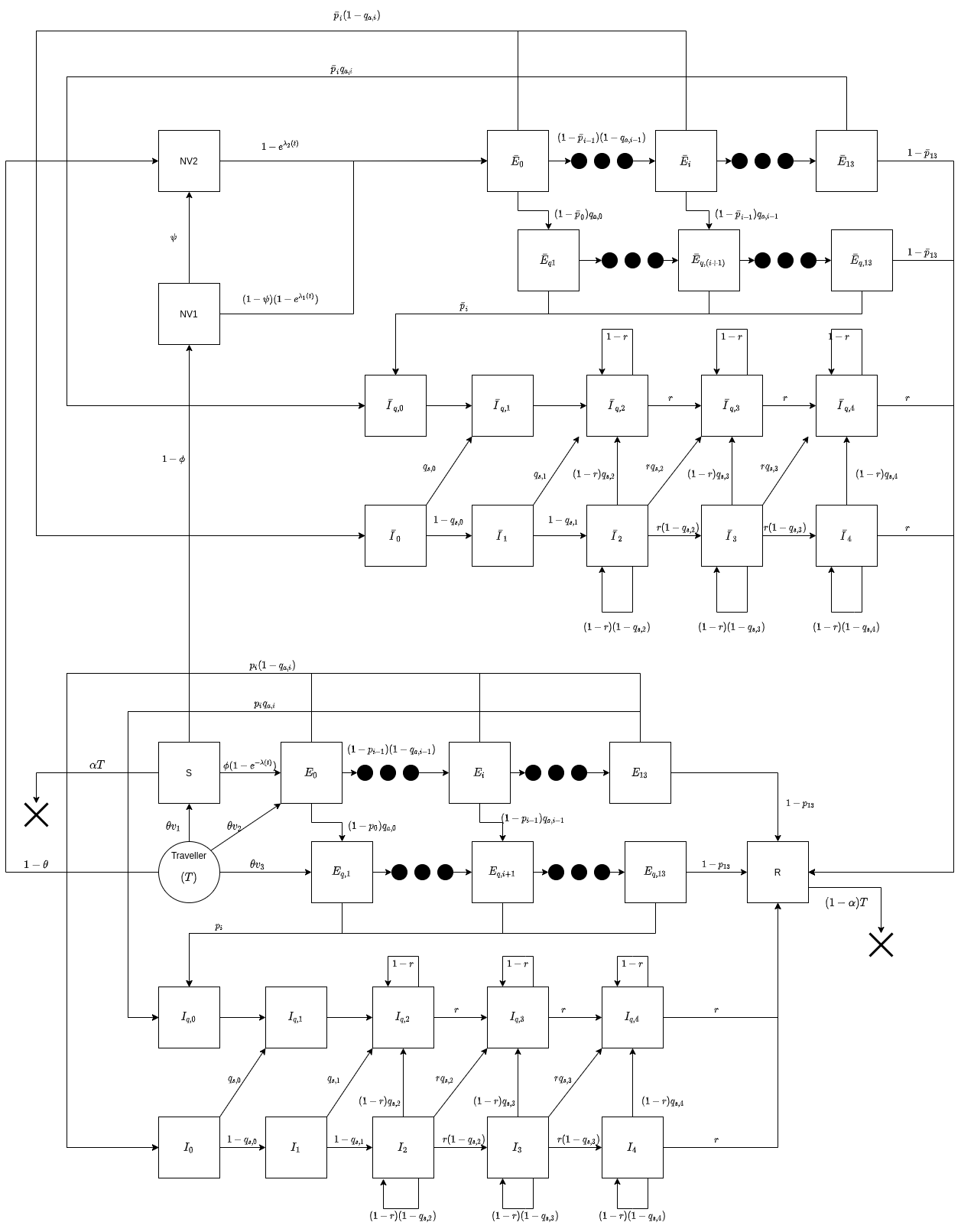}\\
  \caption{Diagram of our compartmental  model including vaccines.}\label{Fig_Seir_Flow_2_base_trans_travel_vaccine}
  \end{center}
\end{figure}

Then in December 2020, the world started vaccinating. For simplicity our model assumes one type of vaccine requiring two doses. Expending to multiple vaccines, possibly some requiring only one dose, follows the same conceptual framework by adding more compartments. To implement vaccination in the EBM model, we introduce new compartments $NV1$, $NV2$, $\bar{E}$, $\bar{E_q}$, $\bar{I}$ and $\bar{I_q}$. While $NV1$ and $NV2$ represent the number of people who have received the first and second dose of vaccination respectively, the $\bar{E}, \bar{I}, \bar{E_{q}}, \bar{I_{q}}$ keeps track of exposures and infections between vaccine doses and post vaccination. They have the exact same sub-structure as the Exposed and Infected compartments used for non-vaccinated groups but parameters might vary. We introduce the new parameter $\psi= 1/\delta$ where $\delta$ is the time gap between doses, as well as $\mu_1$ and $\mu_2$ to account for possible reduced susceptibility of vaccinated individuals. We also assume an 80\% reduction in transmissibility for vaccinated population, which is represented by the parameter $\omega$ which multiplies the contribution of vaccinated compartment to the hazard rate as seen in Equation 37. We assume a 95\% protection as a result of vaccination. This is modeled as $\prod_{i=0}^{13} (1-\bar{p}_i) = 0.95$. This reduces the probability of vaccinated individuals developing symptoms and also reduces the probability of severe infections. The flow diagram with the vaccinated compartments is shown in Fig.~\ref{Fig_Seir_Flow_2_base_trans_travel_vaccine}.

In the EBM, we assume that 100\% of healthcare population are completely vaccinated by December 27, which is when the community starts receiving vaccination. For this, we begin vaccinating Healthcare population on December 22 with an average of 2500 people being vaccinated everyday. The proportion of community population that is vaccinated is based on daily averages from data from \cite{HawaiiDOH} \\
The introduction of new compartments results in our equations being modified to:
\begin{align}
S(t+1) &= e^{-\lambda(t)}S(t) - {\color{red}NV}\\
{\color{red} NV1(t+1)}&{\color{red}=(1-\psi)(1-(1-e^{-\lambda(t)})\mu_1)) NV1(t) +NV}\\
{\color{red}NV2(t+1)}& {\color{red}=\psi (1-(1-e^{-\lambda(t)})\mu_1))NV1(t)+(1-(1-e^{-\lambda(t)})\mu_2))NV2(t)} \\
E_0(t+1) &= (1-e^{-\lambda(t)})S(t)\\
{\color{red}\bar{E}_0(t+1)} &{\color{red}= (1-e^{-\lambda(t)})(\mu_1 NV1(t) + \mu_2 NV2(t))} \\
E_{q,i}(t+1)& = (1-p_{i-1})(q_{a,i-1}E_{i-1}(t)+E_{q,i-1}(t)) \\
{\color{red}\bar{E}_{q,i}(t+1)} &  {\color{red}=(1-\bar{p}_{i-1})(q_{a,i-1}\bar{E}_{i-1}(t)+\bar{E}_{q,i-1}(t))} \\
{\color{red}\bar{I}_{0}(t+1)} & {\color{red}=\sum_{i=0}^{13} \bar{p}_i(1-q_{a,i})\bar{E}_{i}(t)} \\
{\color{red}\bar{I}_{q,0}(t+1)} & {\color{red}=\sum_{i=0}^{13} \bar{p}_i(q_{a,i}\bar{E}_{i}(t)+\bar{E}_{q,i}(t))}
\end{align}
where the terms in red account for vaccination where $NV$ represents the daily average of number of people receiving the first dose of vaccination. With these assumptions and changes in the model, the equation for the hazard rates get modified to: 
\begin{multline}
\lambda_c(t) = \beta(1-p_{mp}(1-p_{me}))\Big[
  (I_c+\varepsilon E_c)+
  \gamma((1-\nu)I_{c,q}+\varepsilon E_{c,q})+\\
  \rho [(I_h+\varepsilon E_h)+  \gamma((1-\nu)I_{h,q}+\varepsilon E_{h,q})]+\\
  {\color{red}\omega((\bar{I}_c+\varepsilon \bar{E}_c)+
  \gamma((1-\nu)\bar{I}_{c,q}+\varepsilon \bar{E}_{c,q}))}+\\
  {\color{red}\omega((\bar{I}_h+\varepsilon \bar{E}_h)+
  \gamma((1-\nu)\bar{I}_{h,q}+\varepsilon \bar{E}_{h,q}))}]\Big]/N_c, \label{vaccine_c_lambda}
\end{multline}
where:
\begin{equation}
    N_c(t) = S_c+E_c+I_c+R_c+\rho(S_h+E_h+I_h+R_h) \\
    {\color{red}+ NV1_c + NV2_c}
\end{equation}
  and $\omega$ is the reduction in transmissibility due to vaccination. If for simplicity we assume that by the start of the vaccination period for community population, the entire healthcare population is fully vaccinated, their hazard rate is modified to:
 \begin{equation}
    \lambda_h(t) = \rho\lambda_c+\beta\eta{\color{red}\omega}\Big[
  (I_h+\varepsilon E_h)+
  \kappa \nu (I_{h,q}+I_{c,q})\Big]/N_h, \label{vaccine_h_lambda}
\end{equation}

When combining travelers and vaccines, we assume that a proportion $\theta$ of all arriving travelers are fully vaccinated. These are moved directly into the NV2 compartment (for respective categories returning resident and tourists). The remaining $1-\theta$ are distributed between Susceptible, Exposed and Exposed quarantine compartment based on our prior assumptions from the model including travelers. 

\begin{figure}[htp!]
\begin{center}
  \includegraphics[width=5in]{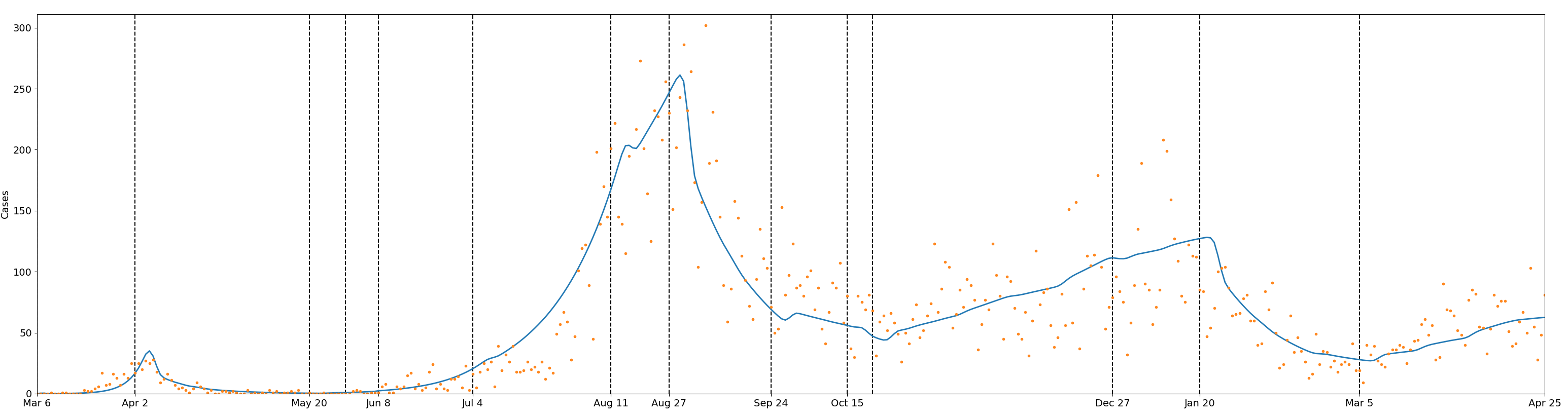}\\
  \includegraphics[width=5in]{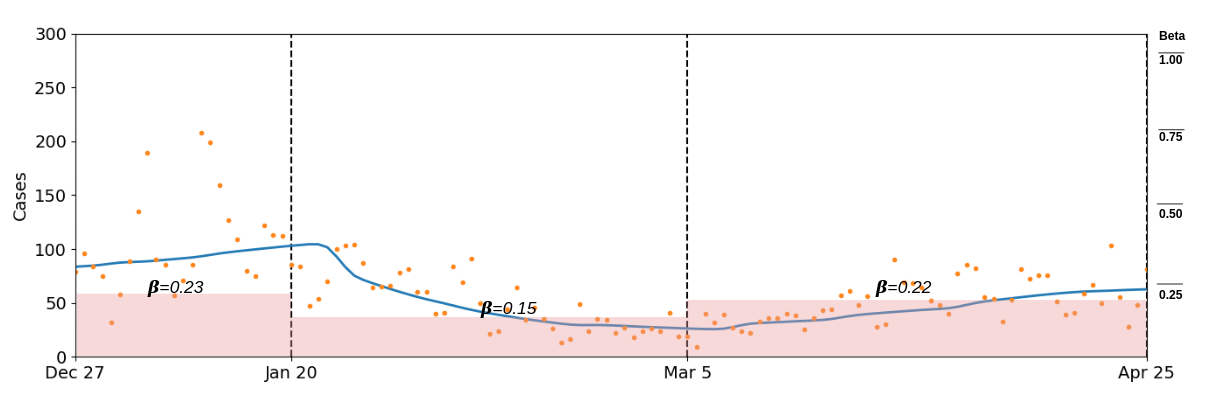}\\
  \caption{Top: Honolulu County fit from March 6, 2020 including travelers starting October 15, 2020 and vaccination starting December 27, 2020. Simulation runs through April 25, 2021. Bottom: Zoom in on period where both vaccination and travelers are included with the corresponding basal transmission rates.}\label{Fig_Seir_EBM_fit_Travelers_Vaccines}
  \end{center}
\end{figure}

Figure \ref{Fig_Seir_EBM_fit_Travelers_Vaccines} including travelers and vaccines. The values of parameters used for Honolulu county for the model including vaccines are shown in Table \ref{tab:parameters-vaccine}. The parameter values and data for travelers is the same as shown in Table \ref{tab:parameters-travelers} and Table \ref{tab:data-traveler} respectively. 
\begin{table}[h!]
\caption{Parameters intrinsic to vaccination used in our simulations}
\centering
\begin{tabular}{ |>{\raggedright\let\newline\\\arraybackslash\hspace{0pt}}p{6.5cm}||>{\raggedright\let\newline\\\arraybackslash\hspace{0pt}}p{6cm}|}
 \hline
\centerline {Parameter, meaning} & \centerline {Value} \\
 \hline
 \hline
  \multicolumn{2}{|c|}{Factors modifying transmission rate}\\
 \hline
$\mu_1$, reduced susceptibility after Dose 1 & 1 (we assume no reduction in susceptibility after dose 1) \\
\hline
$\mu_2$, reduced susceptibility after Dose 2 & 1 (we assume no reduction in susceptibility after dose 2)\\
\hline
$\psi$, fraction of newly fully vaccinated to dose 1 vaccinated & 1/21 \\
\hline
$\omega$, reduced transmissibility due to vaccination & 0.20\\
\hline
$\bar{p}_i$ , $i=0,1,...13$, probability of onset of symptoms after day i (after vaccination) & 0.000492, 0.001080, 0.002056, 0.0415, 0.002376, 0.000858, 0.000528, 0.000302, 0.00019, 0.00019, 0.00019, 0.00019, 0.00019, 0  \\

\hline
$\theta$, proportion of travelers assumed to be unvaccinated & 1\\
\hline 
$NV$, vaccinations received per day & 2500 \\
\hline
\end{tabular}
\label{tab:parameters-vaccine}
\end{table}

\newpage

\subsection{Agent-Based Model}
Compartmental models focus on directly capturing the collective behavior of groups of people, and are typically derived using estimates of aggregate (or limiting) behavior of a large number of individuals under (many) simplifying assumptions. In contrast, agent based models (ABMs) focus on capturing the behavior of a single individual, referred to as an agent. Such individual behavior can often be described using fairly simple rules, however the collective behavior may still exhibit complicated phenomena. As a related example from physics, one could imagine modeling diffusion using the standard PDE versus modeling the Brownian motion of each particle. It should be noted, however, that ABMs have been used in social, economic, and biological sciences as early as the 80s \cite{reynolds1987flocks} , and some models could be tracked to the 70s \cite{schelling1971, schelling1978micromotives}. Popularity of ABMs exploded in the 90s, when the computational power significantly increased and became widely available.

d $S$ can be a product $\mathbb{R}^m\times\mathbb{Z}_2^k$, where $\mathbb{R}^m$ captures continuous variables related to the disease, such as age and susceptibility to infection, and $\mathbb{Z}_2^k$ captures binary variables, such as presence of infection and comorbidities. Of course, one can add other categorical variables if needed.

The evolution of the system state through time can be regarded as a function $g:T\to A^S$ or, equivalently, $f:T\times A\to S$, with $f(t,a)=g(t)(a)$, where $T$ is our time set, which we shall assume to be discrete, i.e. $T=\mathbb{N}$. The function $f$ can be defined deterministically or it can be a realization of a stochastic process. Importantly, one does not focus on the whole $f$ and instead defines how an individual value, $f(t+1, a)$ is obtained when $f(t,a)$ for all $a\in A$ is known. Of course, a change in the state of an agent is unlikely to depend on all of the other agents. Typically, each agent has an associated subset of agents that may affect its state due to ``interaction''. We shall refer to such a subset as a \emph{contact set} of an agent. In the case of pandemic modeling, a contact set consists of people who actually interact with a given individual. This example also suggests that a contact set of an agent $a\in A$, which we shall denote $N(a)$, may have an additional structure to better reflect interactions within different contexts. For example, interactions at work may be different from those at home. Mathematically, this may be represented as a disjoint union, $N(a)=\sqcup_{i=1}^n{N_i(a)}$. Also, a contact set may be time dependent, thus yielding $N(t, a)=\sqcup_{i=1}^n{N_i(t, a)}$, $t\in T$.

Considering that interactions between agents are typically symmetric, it is convenient to represent all $N(a)$, $a\in A$, using an undirected graph, so that each separate $N(a)$ is just a collection of adjacent vertices. More specifically, we let $A$ be the vertex set of our graph, and let $E\subset \{\{a,b\}|a,b\in A\}$ be the set of edges. We shall refer to such a graph as a \emph{contact network}. We can then define a contact set of an agent $a\in A$ as $N(a) = \{b\in A| \exists \{a,b\}\in E\}$. Additional structure and time dependency of contact sets are obtained by defining multiple, possibly time dependent interaction networks with the same vertex set $A$ and different edge sets $E_i(t)$, $i=1,\ldots,n$.

In the deterministic case, the state of an agent $a\in A$ at time step $t+1$ is defined as a function of time $t$ and the states $f(t,b)$ where $b$ ranges over $N_i(t)$, $i=1,\ldots,n$. In the stochastic case, one computes the conditional probability distribution for $f(t+1, a)$, conditioned on the above $f(t,a)$ (and possibly $t$), and then samples from it. When modeling a pandemic, this often simplifies to computing the probability of infection given an uninfected individual, or computing the probability of developing symptoms, given an infected but asymptomatic individual, etc. Then a (pseudo)random number is generated to determine the actual state transition (e.g. an individual gets infected, develops symptoms, etc). While the transition between states can be mathematically quite complicated and non-Markovian, the actual implementation is often fairly straightforward.

\begin{figure}[h!]
\centering
    \begin{subfigure}{0.47\textwidth}
        \includegraphics[width=\textwidth]{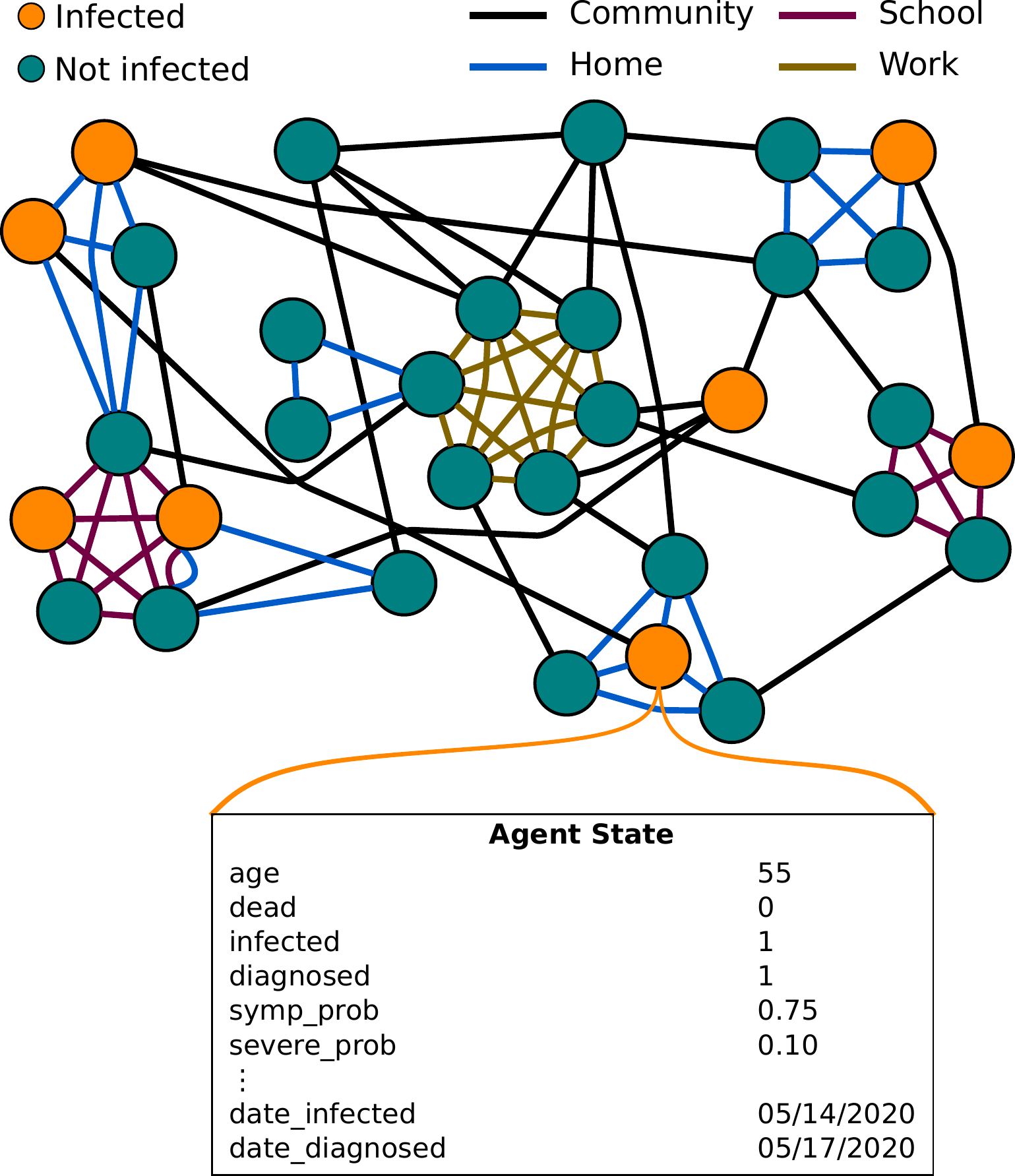}\\
    \end{subfigure}
    \hfill
    \begin{subfigure}{0.47\textwidth}
        \includegraphics[width=\textwidth]{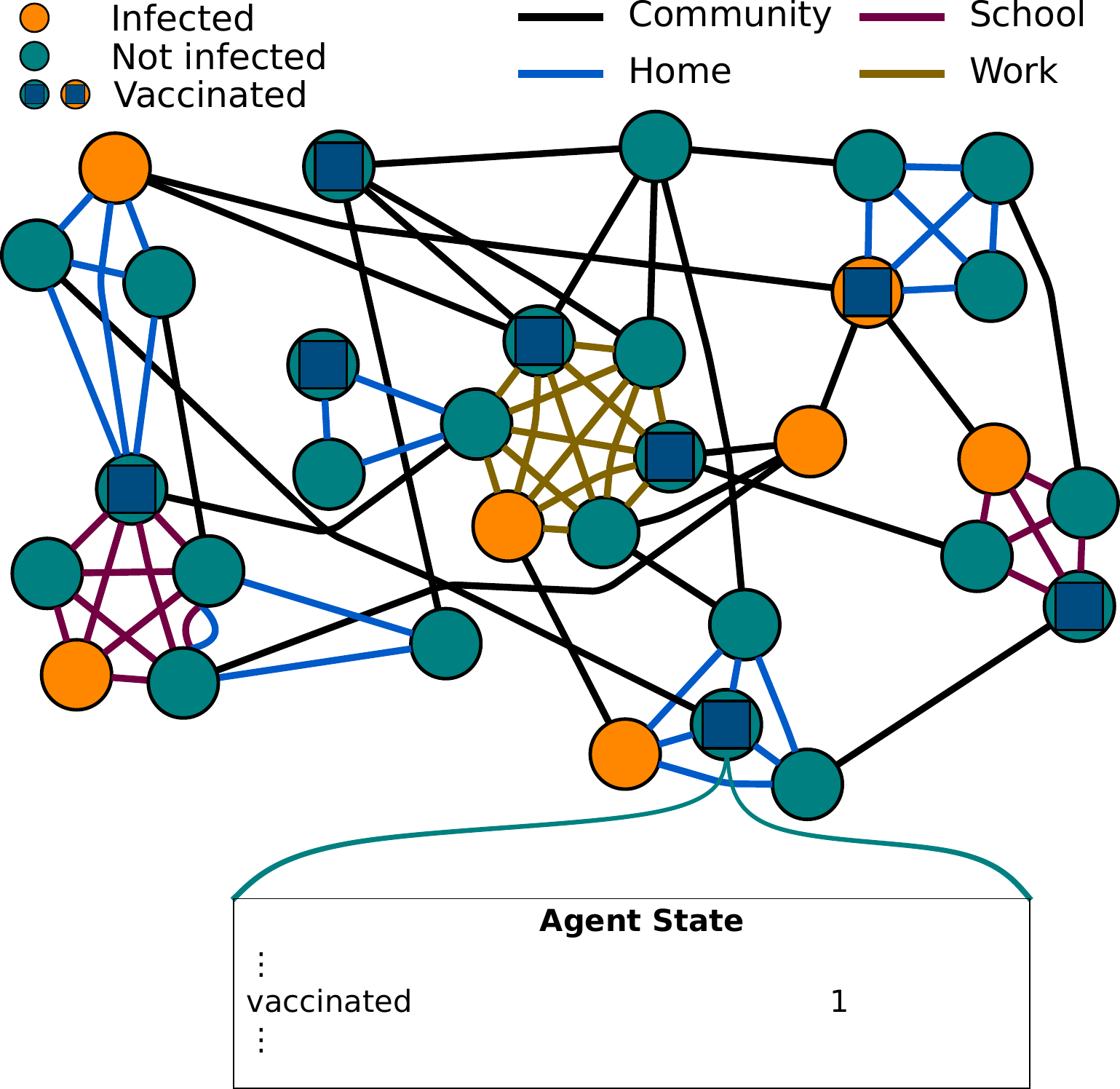}\\
    \end{subfigure}
  \caption{Sample  contact network representing individuals in the population as nodes and the interactions for possible viral transmission among them as edges. The different colors refer to four different types of contacts or individuals in the population (LHS). The vaccinated individuals will have a reduced transmission which is reflected in their state (RHS).} 
  \label{fig:cn}
\end{figure}

An example of a contact network for an agent based model of a pandemic is shown in Fig.~\ref{fig:cn}. We should note that contact networks constitute one of the most important aspects of agent based models and their construction can be a challenging problem. But once the network construction is done, handling changes to agents' states due to newly available information can be readily implemented as tweaks to the interaction network and/or states of agents. For example, Fig.~\ref{fig:cn} shows that adding vaccinated individuals is conceptually quite simple. Of course, the process governing state transitions also needs to be updated.

\subsubsection{Covasim}
For our simulations, we use the open source COVID-19  Agent-based  Simulator (Covasim) \cite{Kerr2020.05.10.20097469}. Covasim provides several ways to construct contact networks, with the default choice resulting in a network consisting of four ``social layers'' (i.e. edge sets) analogous to the ones shown in Fig.~\ref{fig:cn}. The state of a Covasim agent consists of $39$ variables which include demographic information, individual susceptibility, and variables representing intrahost viral dynamics (along with viral-load-based transmissibility). By design, Covasim is a stochastic agent-based model. Thus, as mentioned above, each step is focused on computing multiple transition probabilities.

While Covasim comes with its own demographic data set, it can also incorporate user-supplied demographic information, and we chose to use the data from the \Hawaii{} Population Model developed by the \Hawaii{} Data Collaborative \cite{HawaiiPopu}. In fact, Covasim allows the user to easily customize multiple aspects of the initialization step. We customized the population size along with the number of initially infected people as well as multiple parameters affecting the simulation, such as probability of asymptomatic infection, probability of isolation upon the onset of symptoms, etc. Since Covasim is a stochastic model, output quantities of interest (e.g. new daily infections, the number of hospitalizations, etc.) are averaged over multiple simulations, typically 15 to 20. A user can supply a seed for the (pseudo)random number generator, typically keeping it fixed during the testing phase and then properly resetting the seed before each simulation. In most cases we employed the default construction of the contact network with its four social layers: household, work, school, and community. However, it is fairly straightforward to alter the default construction, for example by increasing the average number of contacts among young adults in the community layer. We actually did implement the latter to better capture the fact that young people are typically more socially active.

Customization of Covasim simulation steps is done through so called ``interventions,'' which are procedures that can be supplied to and then called by the main simulation routine. Covasim comes pre-packages with several interventions, allowing the user to take into account changes in the baseline transmission rate due to imposed mitigation measures, such as a lockdown, as well as incorporate testing and contact tracing with customized isolation probabilities. One can also write custom interventions, which we did to implement the \Hawaii{} vaccination protocol. The latter was necessary because the vaccine intervention bundled with Covasim could not properly capture the necessary dynamics of administering available COVID-19 vaccines.

Once Covasim runs its initialization step, it iterates over the given number of days and for each iteration uses the constructed contact network along with the provided parameters and interventions to calculate the probability that an agent gets infected. Additionally, the state of each already infected agent gets updated (e.g. switching from an asymptomatic infection to a symptomatic one) according to the prognoses pre-calculated during the initialization step.
\begin{figure}[htp!]
\begin{center}
  \includegraphics[width=4in]{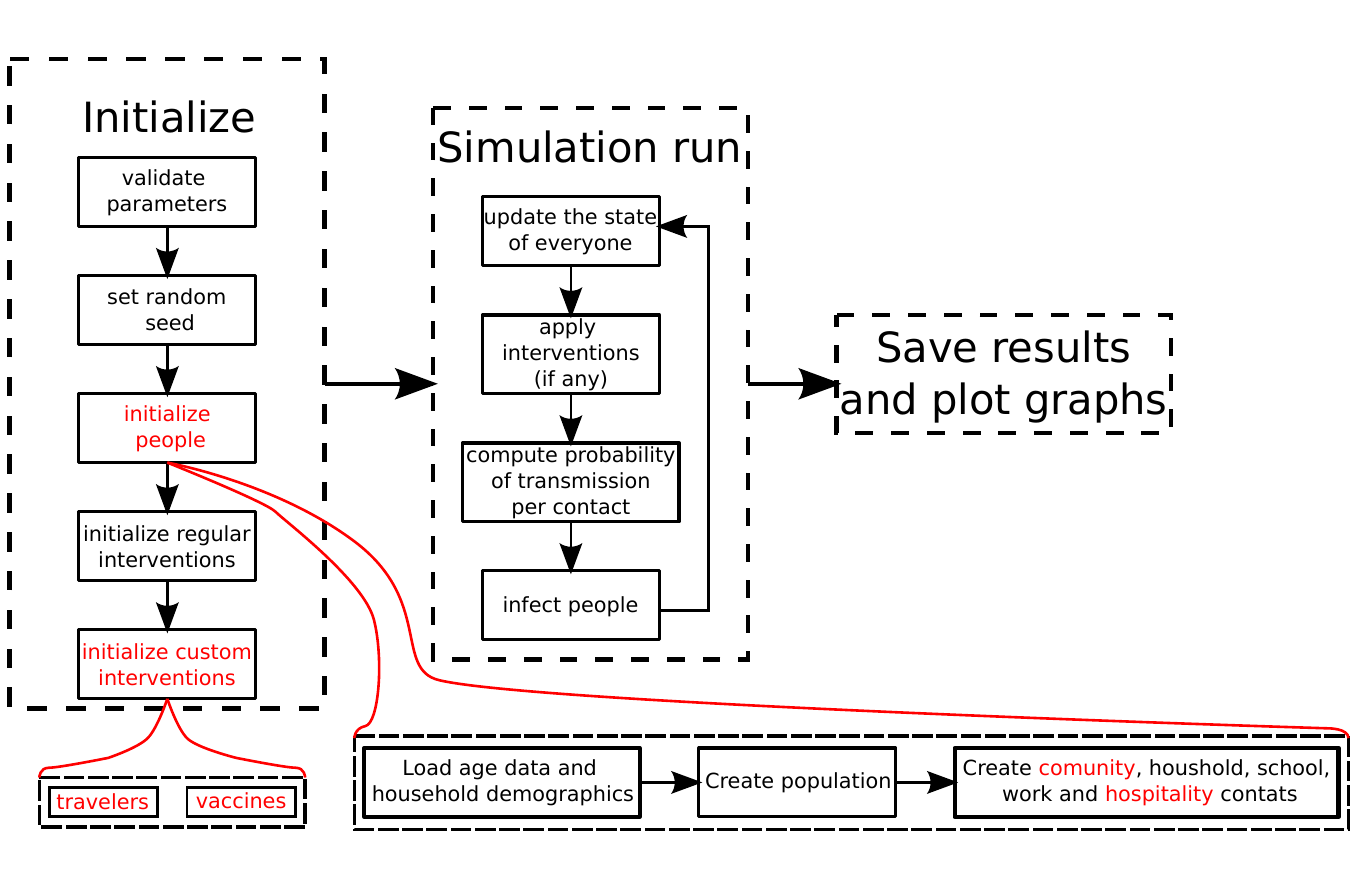}\\
  \caption{Diagram of basic Covasim simulation algorithm.}\label{Fig_Covasim_Diagram}
  \end{center}
\end{figure}

Incorporating the effect of travelers on the spread of the disease is important for \Hawaii{}, but it  had to be done in an indirect way, since Covasim keeps the total number of agents fixed throughout the simulation. Hence, we added a fifth social layer of contacts representing workers in the hospitality industry and a new custom intervention which increased the baseline transmission rate within the new layer based on the number of tourists traveling to \Hawaii{} as well as the nationwide infection rates.

\begin{figure}[htp!]
\begin{center}
  \includegraphics[width=5.2in]{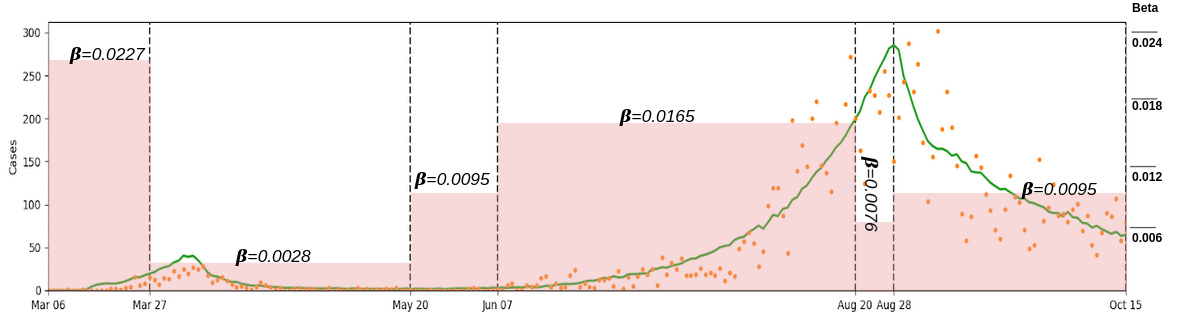}\\
  \caption{In green is the Covasim model fit for Honolulu county from March 6 to October 15, 2020. Included are the optimized transmission rates. }\label{Fig_Seir_ABM_fit_Basic}
  \end{center}
\end{figure}

Figure \ref{Fig_Seir_ABM_fit_Basic} represents the optimized fit obtained from the Covasim model for Honolulu County. 
\begin{figure}[htp!]
\begin{center}
  \includegraphics[width=5.2in]{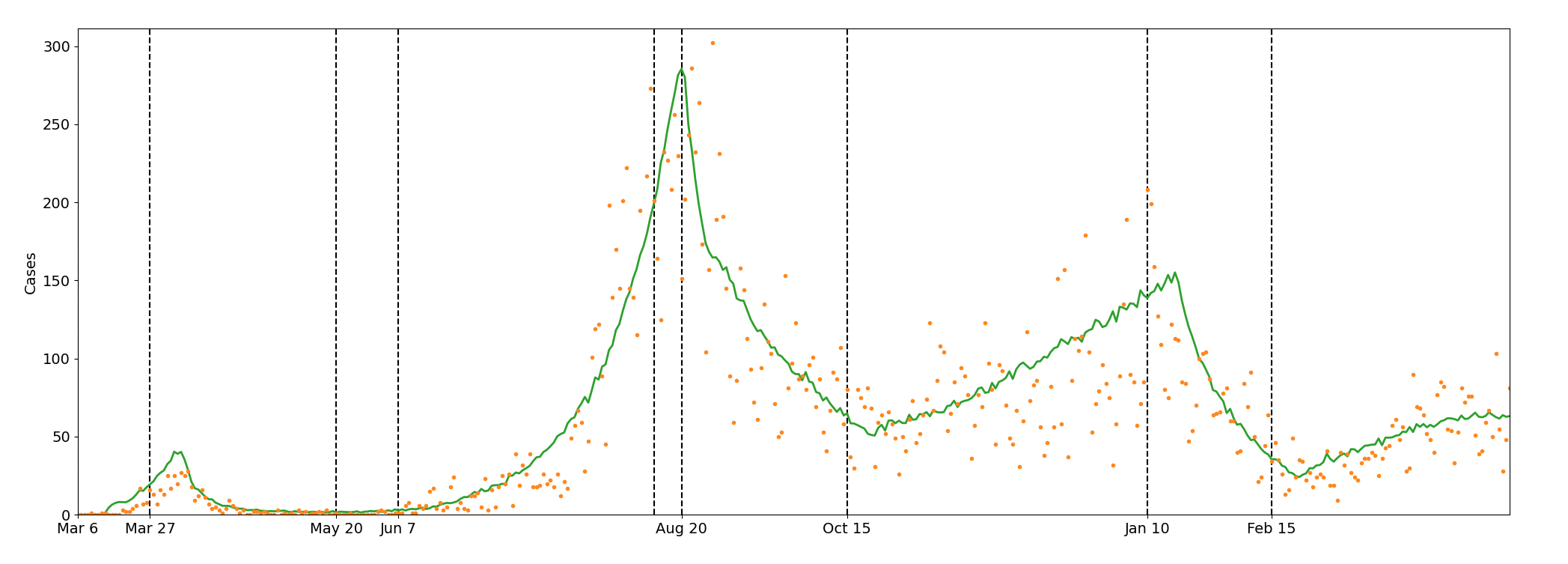}\\
  \includegraphics[width=5.2in]{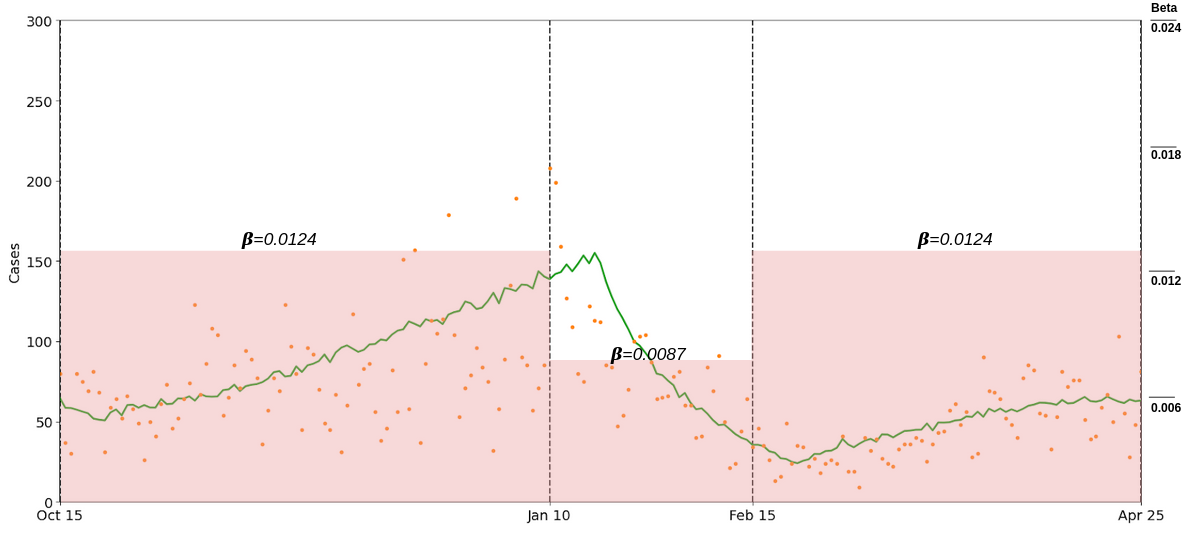}\\
  \caption{Top: Covasim fit for Honolulu county from March 6 to April 25, 2021 with travelers and vaccine included. Bottom: zoom in on the period October 15, 2020 - April 25, 2021 including the corresponding transmission rates.}\label{Fig_ABM_Traveler_Vaccine}
  \end{center}
\end{figure}
It is important to note that while both SEIR and Covasim models denote by $\beta$ the parameter representing the baseline transmission rate of the disease, these two parameters are, in fact, different quantities. In Covasim, it can be regarded as the probability of a susceptible person getting infected when a contact with an infected individual occurs and no information about other modifiers is given (i.e. layer-specific transmission rate factor, individual transmissibility, and individual susceptibility are all equal to $1$). In the standard discrete SEIR model, this parameter also affects the fraction of newly infected individuals, with the latter given by $1-\exp{(-\beta\frac{I}{N})}$, but its relation to the probability of infection per contact is indirect and depends on multiple simplifying assumptions. This relation becomes even more complicated in our expanded SEIR models.
\section{Computational Versus Conceptual Complexity, and Data Fitting}
\label{section-advlimmodels}
In this section we discuss the advantages and limitations of both models. 
\subsection{Conceptual Complexity}
 While computationally efficient, compartmental models present another difficulty: if the model itself needs to be modified to take into account newly discovered features of the pandemic, such modifications can be highly non-trivial. The reason for the difficulty is the aggregate nature of the interactions. Some of these interactions are quite intricate and are based on a series of complicated assumptions. Consequently, the conceptual  complexity of the model may become a hurdle. This can be observed from the diagrams in figures \ref{Fig_Seir_Flow_2_base_trans}, \ref{Fig_Seir_Flow_2_base_trans_travel} and \ref{Fig_Seir_EBM_fit_Travelers_Vaccines} where we illustrate the incorporation of additional compartments into the model. Adding new variants to the compartmental model would drastically increase the already existing complexity of the interactions between different compartments. A compartmental model is also quite limiting when it comes to the inclusion of demographic, ethnic and other essential information, as it would again require one to introduce numerous compartments. 

 For agent based models, incorporating new attributes for individuals, such as age, ethnicity or vaccination status, is conceptually fairly simple, since each individual is represented by an agent. Thus, a simple augmentation of the variables representing the state of each agent should be done, and the network remains unchanged. For instance, taking into account a new variant of SARS-CoV-2 would amount to simply adding variables that characterize the variant to the agent state space.
 \subsection{Computational Complexity}
 Simulation of agent based models may be a computationally expensive  process if the number of agents (i.e., the population) is large and the computation is not appropriately parallelized.  Each time step requires an iteration over each interaction edge, and the state of each agent needs to be evaluated and possibly updated. Thus, the total time complexity of each time step is $O(|E|+|A|)$. Here $|E|$ denotes the number of interaction edges and $|A|$ denotes the number of agents. Multiplying this already substantial number by the number of time steps, $N$, results in a fairly large computational cost $O(N(|E|+|A|))$. We should also mention that interventions may also increase the time complexity. Fortunately, in many cases such computations can be parallelized, and employing compute clusters and/or multicore nodes can provide a significant speed-up. However, this potential parallelism is not currently exploited in our simulations, although the 15-20 simulations needed to average the results (as mentioned earlier) are run parallel. 
 
\begin{figure}[htp!]
\begin{center}
  \includegraphics[width=4.0in]{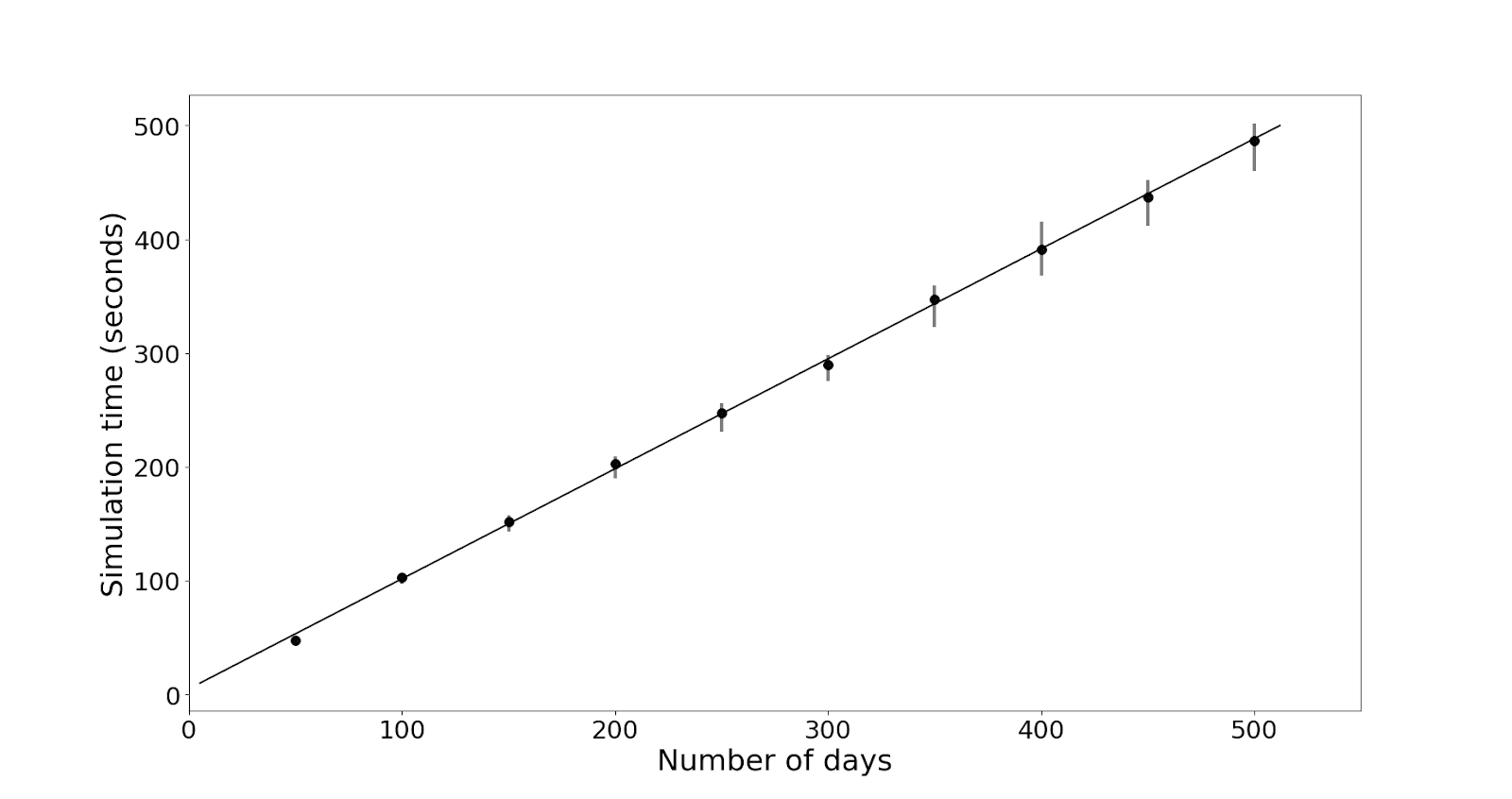}\\
  \includegraphics[width=4.0in]{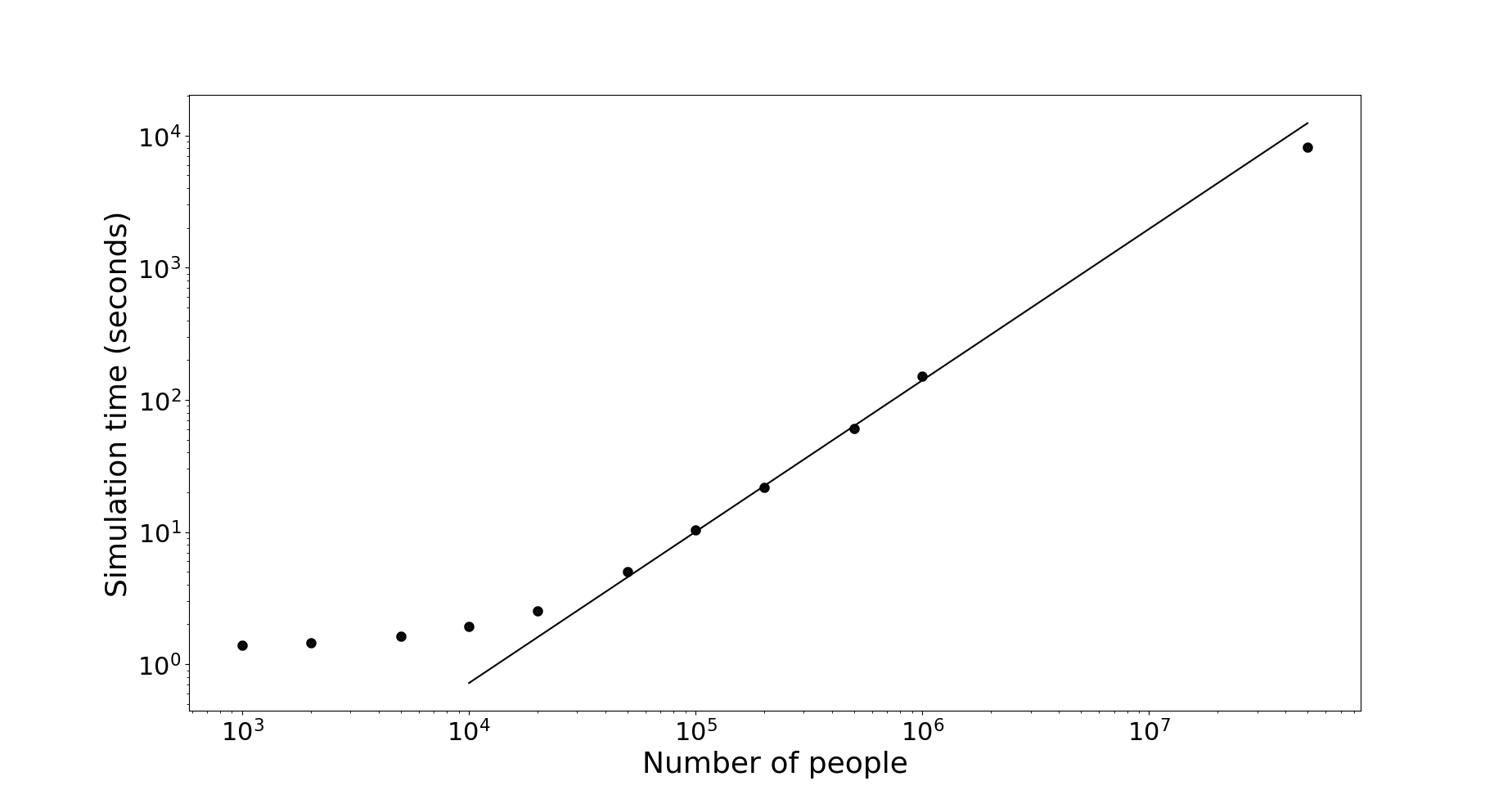}\\
  \caption{Increase in simulation time as a function of problem size for serial Covasim runs. For the top graph, we set the population size equal to $1$ million and run 15 simulations for each scenario with a different number of days starting from 50 to 500. For the bottom graph one we set the number of days equal to 150 and run 15 simulations for each scenario with different population sizes, ranging from a thousand to a million, with one additional point for a population of size $50$ million.}\label{scalecovasim}
  \end{center}
\end{figure}

If $|E|$ is proportional to $|A|$, then one can expect a linear increase in computational time with respect to the number of days and the number of agents.  Figure \ref{scalecovasim} shows this for our simulations. Here, no extra parallelism is employed. These are run on a single Haswell node of the NERSC Cori system. Each node has two sockets, and each socket is populated with a 2.3 GHz 16-core Haswell processor (Intel Xeon Processor E5-2698 v3).
Each core supports 2 hyper-threads, and has two 256-bit-wide vector units. To produce the results in the picture, we used Covasim without additional parallelism and performed the computations on a single core. One can see that the increase in computational time with the number of days is roughly linear, and a reference straight line is included (top figure). The scaling of the computational time with respect to the number of agents is also close to being linear (bottom figure).
In the future, we plan to investigate the role of parallelism in reducing the computational time of agent based epidemiological models. Possible speedup is crucial for large population sizes since, as shown in Fig.~\ref{scalecovasim}, already for 50 million agents one has to wait more than 2 hours to perform required computations on a single node. Moreover, larger populations may require employing distributed computations, as the total state of the model may not fit into the RAM of a single node. 

\begin{figure}[htp!]
\begin{center}
  \includegraphics[width=4.2in]{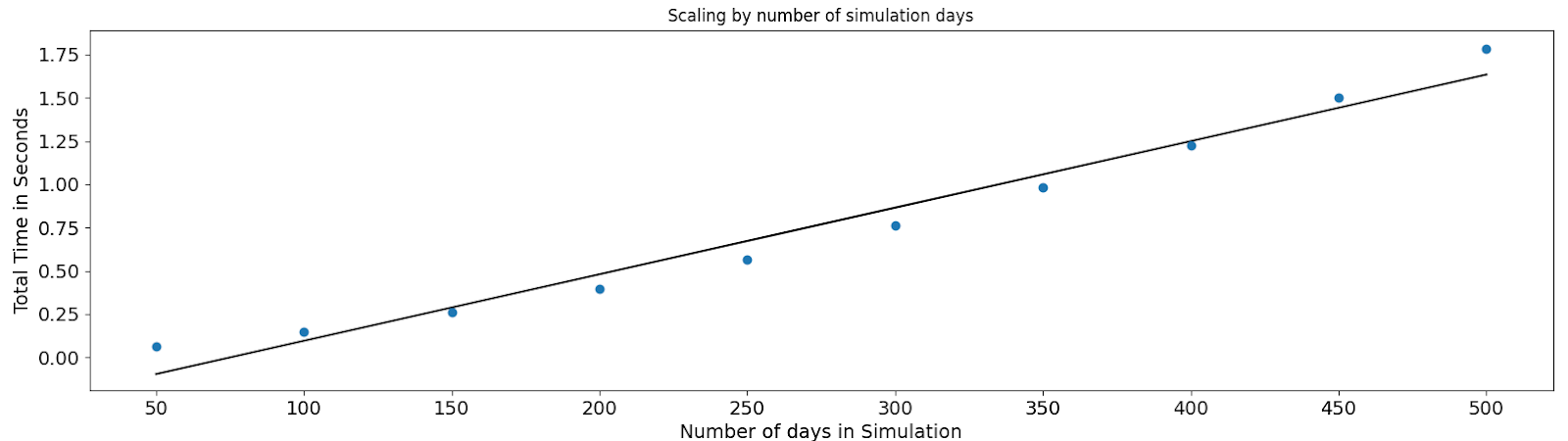}\\
  \caption{Increasing computational cost of the compartmental model. For each number of simulation days, an average of 15 simulations is taken. }\label{scaleSEIR}
  \end{center}
\end{figure}

 Computational complexity of compartmental models is typically $O(NC)$, where $NC$ is the number of compartments, which is significantly less computationally intensive for the same problem. As expected, Fig.~\ref{scaleSEIR} shows a linear increase in the computational time with respect to the number of days. For this benchmark, a basic model with community and healthcare compartments is used without any interventions, travelers, or vaccinations. It was run on single Haswell node of the NERSC Cori system. Note that scaling with respect to the population is not relevant for a compartmental model since the population is aggregated. Generally agent based models are likely easier to parallelize, however we have not explored this because of the Python-based code in Covasim and its reliance on Python specific libraries which make hands-on parallelization techniques on a node,  such as OpenMP, more difficult. 
\subsection{Data Fitting}
There are differences in data fitting methodology, implementation and results for the two models we study. Our SEIR model was fitted to available data from the State of \Hawaii{} based on daily infections using a classical gradient-based optimization method. The latter is possible because we can explicitly compute the gradient with respect to the parameters. Thus obtained values of the baseline transmission rate for the SEIR model allowed us to calculate appropriate values of the corresponding parameter in Covasim. Fitting an agent based model to data directly is a far more complicated task. In most cases, one needs to resort to a very general global optimization technique such as the Metropolis-Hastings algorithm or the genetic algorithm \cite{chib1995, hendrix2010}. Such a procedure is very computationally costly and does not always produce a good fit \cite{hendrix2010}. 

\begin{figure}[htp!]
\begin{center}
  \includegraphics[width=5.2in]{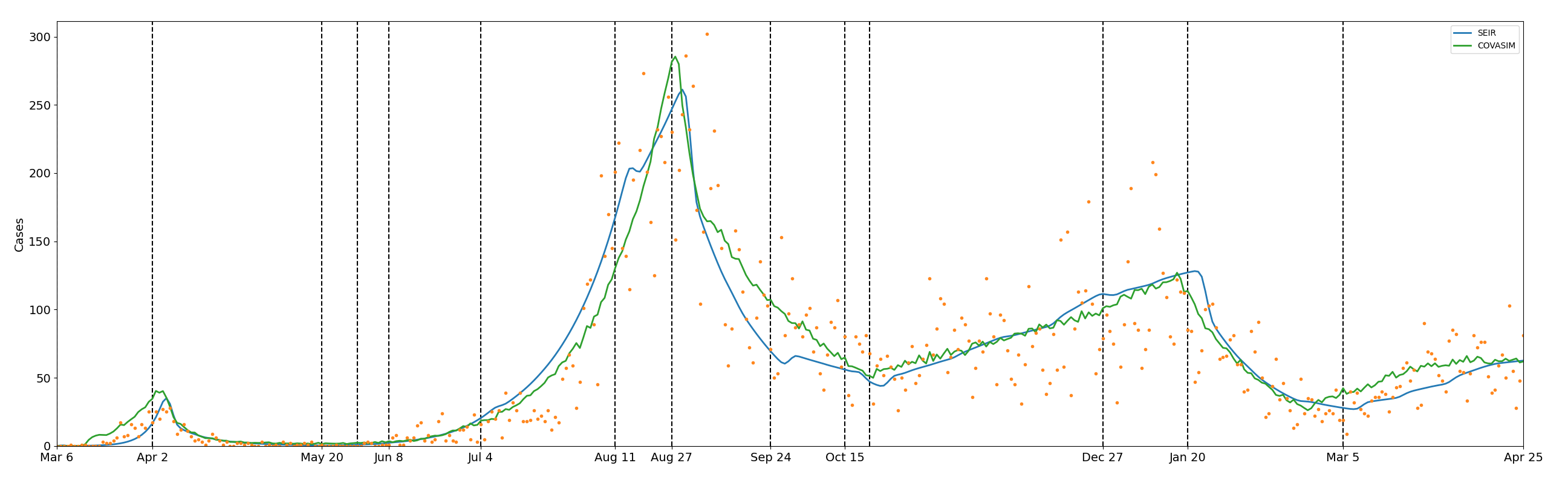}\\
  \caption{Compartmental(blue) and Covasim (green) fit from March 6, 2020 to April 25, 2021 including travelers and vaccines. }\label{Fig_BothFit}
  \end{center}
\end{figure}

Figure \ref{Fig_BothFit} shows the optimized fit obtained from the SEIR and the Covasim model for Honolulu County. We can see that qualitatively both fits agree very well in certain places, including the spike in August. 

\begin{figure}[htp!]
\begin{center}
  \includegraphics[width=4.2in]{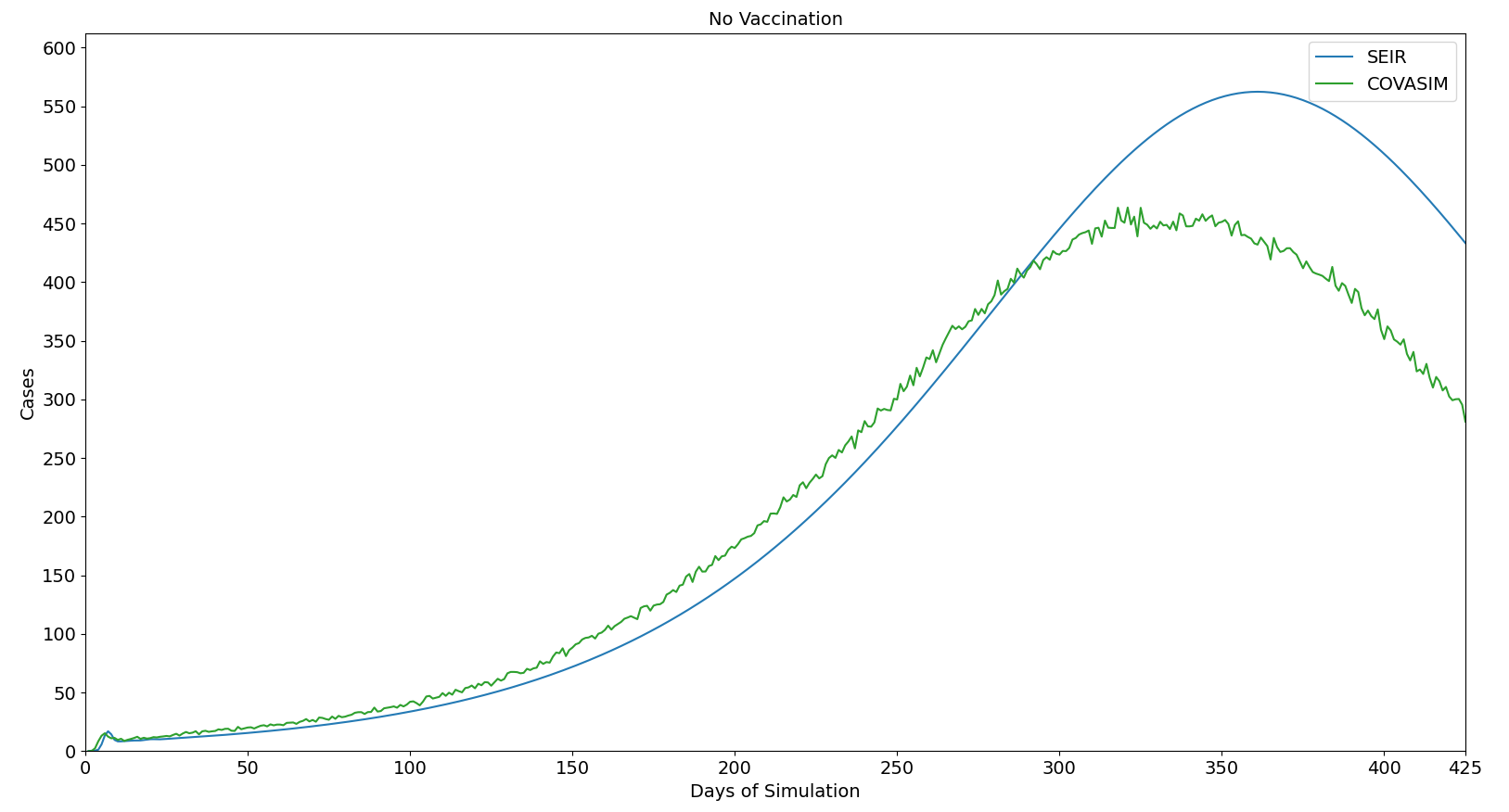}\\
  \includegraphics[width=4.2in]{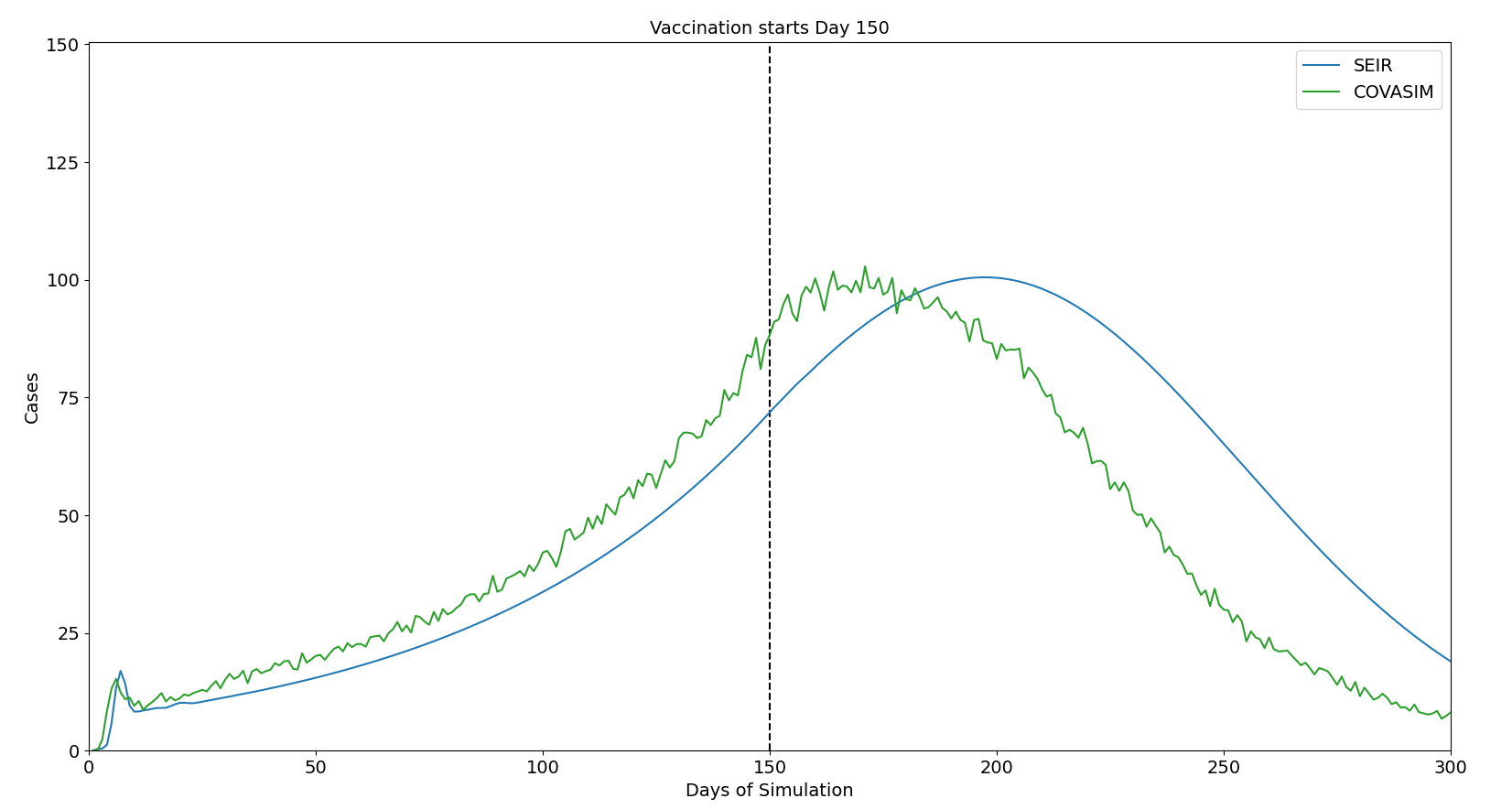}\\
  \caption{SEIR (blue) and Covasim(green) benchmark scenarios forecasting spread and vaccination.}\label{Fig_Benchvaccines}
  \end{center}
\end{figure}

While the models agree well when fitting specific data as for Honolulu County, it is unclear if the agreement will holds as we run the model without fitting data. In Fig.~\ref{Fig_Benchvaccines} we show simulation of a benchmark scenario to visualize the impact of the vaccines for both models. We start with a basic model with no interventions, vaccinations or travelers. For both models, we assume a population of 1 million, with 100 infections on day 1. In EBM this is accomplished by setting $E_0(0)$=100 for the community population and choosing a single beta value for the simulations. We forecast two scenarios, the first one has no vaccines and shows comparatively how the two models forecast over a year. For the second scenario, we assume 2500 individuals are vaccinated per day, a 95\% vaccine protection for developing symptoms (this means that $\bar p_i$ are chosen such that $\Pi_{i=0}^{13} (1-\bar p_i)=0.95$), and a 80\% reduction in transmissibility (which corresponds to $\omega=0.2$). We also assume the vaccine requires only one shot. We can observe that the model forecasting agrees even with no fitting and overall the models behave similarly.

\begin{figure}[htp!]
\begin{center}
  \includegraphics[width=5.2in]{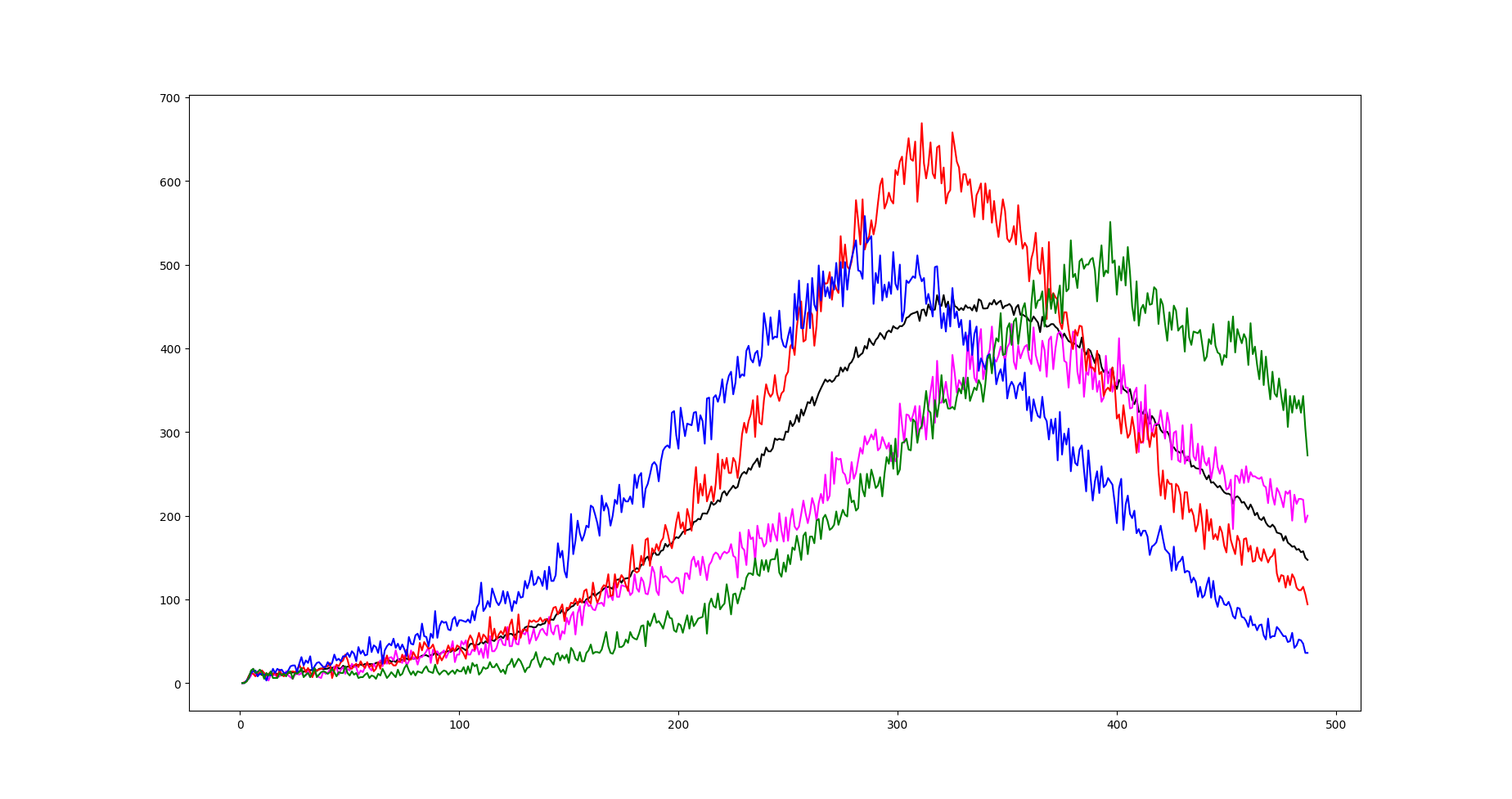}\\
  \includegraphics[width=5.2in]{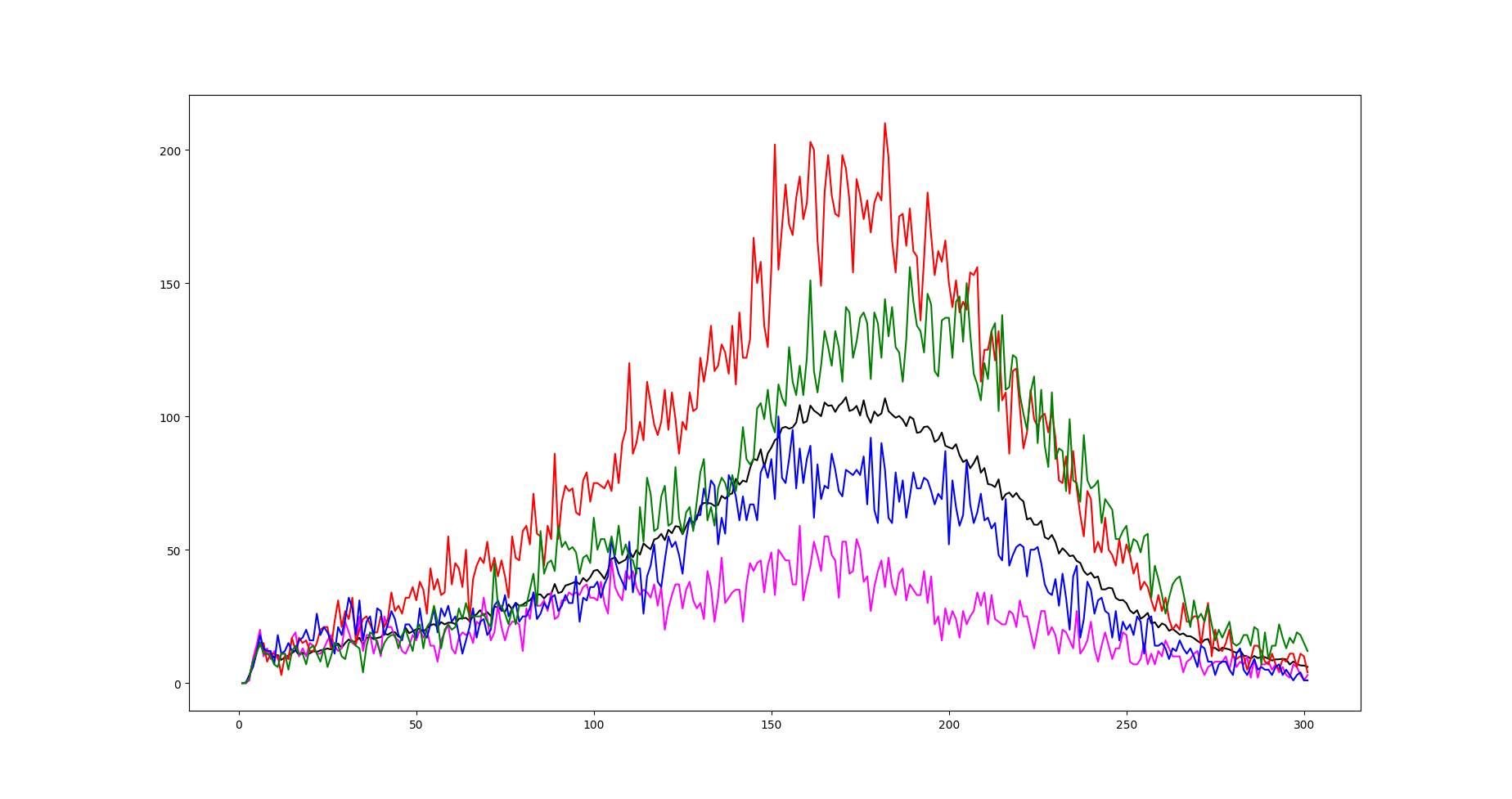}\\
  \caption{The mean of 20 covasim simulations (black) with simulations with highest (red), lowest (magenta), earliest (blue) and latest (green) peaks.}\label{Fig_Mean_Picks}
  \end{center}
\end{figure}

It is important to note that the Covasim curve above is actually the mean of 20 simulation curves. In Fig.~\ref{Fig_Mean_Picks} you can see the mean with 4 individual simulations curves: one with the highest peak (red), one with the lowest peak (magenta), with the earliest peak (blue), and with the latest peak (green). Taking the values (and times of occurrence) of all the peaks of our $20$ simulations and computing their means for the scenario without vaccines, we get the mean peak value of $520$ and the mean time of occurrence of $341.05$, while the value of the peak of the mean of all simulations is $463.6$ occurring at time $321$. For the second scenario, the mean value of the peaks $114.95$ with the mean time of occurrence of $178.15$, while the value of the peak of the mean of the simulations is $102.05$ occurring at time $172$. These observations suggest that the differences in the curves in Fig.~\ref{Fig_Benchvaccines} are at least partially caused by somewhat a simplistic computation of the averages in Covasim.

 \section{Conclusion}
 \label{section-conclusion}
 In this paper, we look at two types of epidemiological models and analyze their complexity both in terms of conceptual design and computational time. Our conclusion is that the decision to use one model versus the other ones depends on the objectives, available data as well as access to resources. If used properly, these two types of models offer similar outcomes for the spread of the disease at the population level. This is not surprising, it was indeed observed in \cite{Ozmen} over data from the 1918 Influenza pandemic. Our example, see Fig.~\ref{Fig_Benchvaccines} is designed to understand whether or not the two models agree when no data fitting is performed. It is clearly the case, and demonstrates that overall the models behave similarly. 
 
 Reflecting upon a year of modeling in the current COVID-19 pandemic, we conclude that for the State of \Hawaii{} both models played an important role. Early in the pandemic the compartment model allowed us to fit the data, and run some forecasting with limited data and mitigation measures that applied to the entire population (such as stay-at-home orders). Once the pandemic advanced and actions got more sophisticated with testing, contact tracing, safe travel program, tier system for reopening strategy and eventually vaccines it was clear that shifting to an agent based model was a better strategy. It allowed us to mimic the age demographic for vaccines plan, and include other attribute in an easier way. Computationally, because of the small size population in the State of \Hawaii{} it was not a major issue to use the agent based model. However for a state like California with almost 40 millions people advanced computational resources are required to run an agent based model.  
 
 Developing an hybrid model with some aspects that are agent based but some also with aggregated population might provide in the end the ideal tool.

\section{Acknowledgments}

This material is based upon work supported by the National Science Foundation under Grant No. 2030789.
This research used resources of the National Energy Research Scientific Computing Center (NERSC), a U.S. Department of Energy Office of Science User Facility located at Lawrence Berkeley National Laboratory, operated under Contract No. DE-AC02-05CH11231.
The authors are grateful to the the developers of Covasim (COVID-19 Agent-based Simulator), an open-source model, for useful discussions on the use of their software and the incorporation of new facets such as tourists and vaccines. Covasim is supported by Bill and Melinda Gates through the Global Good Fund.


\printbibliography


\end{document}


\section{Supporting Information}
The equations for the dynamics of the three population groups are essentially the same and are given below. Only the hazard rate and the parameters determining transition rates into quarantine may be different between the three groups.

\begin{align}
S(t+1) &= e^{-\lambda(t)}S(t)\\
E_0(t+1) &= (1-e^{-\lambda(t)})(S(t)-N_{V})\\
\bar{E}_0(t+1) &= (1-e^{-\lambda(t)})N_{V}\\
N_{V}(t+1) &= N_{V}(t)-\bar{E}_{0}+\#daily\_vaccine \\
E_{i}(t+1)& = (1-p_{i-1})(1-q_{a,i-1})E_{i-1}(t),\nonumber\\
&\quad\quad i=1,\ldots,13\\
\bar{E}_{i}(t+1)& = (1-\bar{p}_{i-1})(1-q_{a,i-1})\bar{E}_{i-1}(t),\nonumber\\
&\quad\quad i=1,\ldots,13\\
E_{q,i}(t+1)& = (1-p_{i-1})(q_{a,i-1}E_{i-1}(t)+\nonumber\\
&\quad +E_{q,i-1}(t)), \quad i=1,\ldots,13\\
\bar{E}_{q,i}(t+1)& = (1-\bar{p}_{i-1})(q_{a,i-1}\bar{E}_{i-1}(t)+\nonumber\\
&\quad +\bar{E}_{q,i-1}(t)), \quad i=1,\ldots,13\\
I_{0}(t+1) & =\sum_{i=0}^{13} p_i(1-q_{a,i})E_{i}(t) \\
\bar{I}_{0}(t+1) & =\sum_{i=0}^{13} \bar{p}_i(1-q_{a,i})\bar{E}_{i}(t)
\end{align}
\begin{align}
I_{1}(t+1) & =(1-q_{s,0})I_{0}(t)\\
\bar{I}_{1}(t+1) & =(1-q_{s,0})\bar{I}_{0}(t)\\
I_{2}(t+1) & =(1-q_{s,1})I_{1}(t)+(1-r)(1-q_{s,2})I_{2}(t)\\
\bar{I}_{2}(t+1) & =(1-q_{s,1})\bar{I}_{1}(t)+(1-r)(1-q_{s,2})\bar{I}_{2}(t)\\
I_{j}(t+1) & =r(1-q_{s,j-1})I_{j-1}(t)+\nonumber\\
&\quad +(1-r)(1-q_{s,j})I_{j}(t),\quad j=3,4\\
\bar{I}_{j}(t+1) & =r(1-q_{s,j-1})\bar{I}_{j-1}(t)+\nonumber\\
&\quad +(1-r)(1-q_{s,j})\bar{I}_{j}(t),\quad j=3,4\\
I_{q,0}(t+1) & =\sum_{i=0}^{13} p_i(q_{a,i}E_{i}(t)+E_{q,i}(t))\\
\bar{I}_{q,0}(t+1) & =\sum_{i=0}^{13} \bar{p}_i(q_{a,i}\bar{E}_{i}(t)+\bar{E}_{q,i}(t))\\
I_{q,1}(t+1) & = I_{q,0}(t)+q_{s,0}I_{0}(t)\\
\end{align}
\begin{align}
\bar{I}_{q,1}(t+1) & = \bar{I}_{q,0}(t)+q_{s,0}\bar{I}_{0}(t)\\
I_{q,2}(t+1) & = I_{q,1}(t)+q_{s,1}I_{1}(t)+\nonumber\\
&\quad +(1-r)(q_{s,2}I_{2}(t)+I_{q,2}(t))\\
\bar{I}_{q,2}(t+1) & = \bar{I}_{q,1}(t)+q_{s,1}\bar{I}_{1}(t)+\nonumber\\
&\quad +(1-r)(q_{s,2}\bar{I}_{2}(t)+\bar{I}_{q,2}(t))\\
I_{q,j}(t+1) & = r(q_{s,j-1}I_{j-1}(t)+I_{q, j-1}(t))+\nonumber\\
&\quad +(1-r)(q_{s,j}I_{j}(t)+I_{q,j}(t)),\quad j=3,4\\
\bar{I}_{q,j}(t+1) & = r(q_{s,j-1}\bar{I}_{j-1}(t)+\bar{I}_{q, j-1}(t))+\nonumber\\
&\quad +(1-r)(q_{s,j}\bar{I}_{j}(t)+\bar{I}_{q,j}(t)),\quad j=3,4\\
R(t+1) &= R(t) + rI_{4}(t)+rI_{q,4}(t) + \nonumber\\
&\quad +(1-p_{13})E_{13}(t) + (1-p_{13})E_{q,13}(t) + \nonumber\\
&\quad +r\bar{I}_{4}(t)+r\bar{I}_{q,4}(t)+(1-\bar{p}_{13})\bar{E}_{13}(t) + \nonumber\\
&\quad +(1-\bar{p}_{13})(\bar{E}_{q,13}(t)
\end{align}
\begin{itemize}
 \item\textbf{Variable $S(t)$.} The number of total susceptible individuals.\\
 \item\textbf{Variable $N_V$.} The number of vaccinated susceptible individuals.\\
 \item\textbf{Variables $E_{i}(t)$.} The number of asymptomatic infected individuals $i$ days after exposure who are not quarantined.\\
 \item\textbf{Variables $\bar{E}_{i}(t)$.} The number of vaccinated asymptomatic infected individuals $i$ days after exposure who are not quarantined.\\
 \item \textbf{Variables $E_{q,i}(t)$.} The number of quarantined asymptomatic infected individuals $i$ days after exposure.\\
  \item \textbf{Variables $\bar{E}_{q,i}(t)$.} The number of vaccinated quarantined asymptomatic infected individuals $i$ days after exposure.\\
 \item \textbf{Variables $I_{j}(t)$, $i=0,1$.} The number of symptomatic infected individuals $i$ days after the onset of symptoms who are not quarantined.\\
 \item \textbf{Variables $\bar{I}_{j}(t)$, $i=0,1$.} The number of vaccinated symptomatic infected individuals $i$ days after the onset of symptoms who are not quarantined.\\
 \item \textbf{Variables $I_{j}(t)$, $j=3,4,5$.} The number of symptomatic infected individuals at the nominal stage $i$ of the illness. Note that a person can stay at a given stage for several days.\\
  \item \textbf{Variables $\bar{I}_{j}(t)$, $j=3,4,5$.} The number of vaccinated symptomatic infected individuals at the nominal stage $i$ of the illness. Note that a person can stay at a given stage for several days.\\
 \item \textbf{Variables $I_{q,j}(t)$, $j=0,1$.} The number of quarantined symptomatic infected individuals, with $j$ representing either the number of days after the onset of the symptoms ($j=0,1$), or the stage of the illness ($j=2,3,4$).\\
 \item \textbf{Variables $\bar{I}_{q,j}(t)$, $j=0,1$.} The number of vaccinated quarantined symptomatic infected individuals, with $j$ representing either the number of days after the onset of the symptoms ($j=0,1$), or the stage of the illness ($j=2,3,4$).\\
 \item \textbf{Variable $R(t)$.} The number of removed (recovered or deceased) individuals.
 \end{itemize}

Splitting exposed individuals into multiple stages, $E_{i}$, allows us to capture possible differences in the progression of the asymptomatic phase of the disease. Importantly, it allows us to take into account that, according to the Centers for Disease Control and Prevention (CDC) as well as other sources, about 40\% of people who contract SARS-CoV-2 remain asymptomatic, and the incubation period for those who do develop symptoms is somewhere between 2 to 14 days after exposure, with the mean incubation period between 4 and 6 days \cite{park2020systematic,aoim,pnas}. Individuals who do not develop symptoms after 14 days are assumed recovered. The use of the quarantine sub-compartments, $E_{q,i}$, allows us to capture the effect of contact tracing and the reduced transmission rate for quarantined individuals.

Similarly, having multiple stages for infected individuals better reflects progression of the symptomatic phase of the disease. The first two stages represent the first two days of symptoms, but the next three should be understood as phases of the immune system fighting the disease. There is a substantial variability (due to age as well as other factors) in the number of days any given person can spend at each stage. Our model implicitly assumes that the symptomatic phase of the illness lasts at least 5 days (in the unlikely case that each stage lasts just one day). 
 
 As we mentioned, a crucial part of the dynamics relates to the hazard rate. For the general community, group C, we have

\begin{multline}
\lambda_c(t) = \beta(1-p_{mp}(1-p_{me}))\Big[
  (I_c+\varepsilon E_c)+
  \gamma((1-\nu)I_{c,q}+\varepsilon E_{c,q})+\\
  0.33(\bar{I}_c+\varepsilon \bar{E}_c)+
  \gamma((1-\nu)\bar{I}_{c,q}+\varepsilon \bar{E}_{c,q})+ \\
  \rho[(I_h+\varepsilon E_h)+  \gamma((1-\nu)I_{h,q}+\varepsilon E_{h,q})]+\\
   \rho_{v}[(I_v+\varepsilon E_v)+  \gamma((1-\nu)I_{v,q}+\varepsilon E_{v,q})]\Big]/(N_c+\rho_{v}N_v),
\end{multline}
 and for the tourists we have
\begin{multline}
\lambda_v(t) =\frac{ \rho_{v} \beta\lambda_c+ 
  \beta_v(1-p_{mp}(1-p_{me}))\Big[
  (I_v+\varepsilon E_v)+
  \gamma((1-\nu)I_{v,q}+\varepsilon E_{v,q})\Big]}{(\rho_{v}N_c+N_v)},
\end{multline}
where we suppressed the dependency on $t$ on the right for convenience. We use sub-indices $c$ (community), $h$ (healthcare workers), and $v$ (tourists) to indicate the appropriate group. Subscript $q$ indicates quarantined individuals. Here $p_{me}$ and $p_{mp}$ represent mask efficiency and mask compliance. Mask efficiency is chosen to reflect a reduction in transmission of $75\%$ for all regions. Mask compliance is set at $20\%$ for all regions at the start of the pandemic, but this value is modified on the dates the regions introduce mask regulations. $N_{v}$ denotes the mixing pool for the visitors and $N_c$ denotes the mixing pool for the general community, computed as
\begin{equation}
    N_c(t) = S_c+E_c+I_c+R_c+\rho(S_h+E_h+I_h+R_h)+\rho_{v1}(S_v+E_v+I_v+R_v).
\end{equation}
where variables $E$ and $I$ here represent the sum over all the stages within these compartments. 
For the healthcare worker group, we have
\begin{equation}
    \lambda_h(t) = \rho\lambda_c+\beta\eta\Big[
  (I_h+\varepsilon E_h)+
  \kappa\nu(I_{h,q}+I_{c,q}+I_{v,q})\Big]/N_h,
\end{equation}
where $N_h(t) = S_h+E_h+I_h+R_h$. 

The model fit plot the following value
\begin{equation}
    \label{eq-fitplot}
    \sum_{x=c,h,v}\Big(\sum_{i=1}^3 q_{s,i}^xI(i)+(1-r) q_{s,4}^xI(4)+\sum_{i=1}^12 q_{a,i}^xE(i) \Big)
\end{equation}